\documentclass[%
amsmath,amssymb,
aps,
pre,
twocolumn,
superscriptaddress
]{revtex4-2}

\usepackage{amsmath,amsfonts}

\usepackage{graphics}
\usepackage{graphicx}
\usepackage{dcolumn}
\usepackage{bm}
\usepackage{geometry}
\usepackage{url}
\usepackage{subfigure}
\usepackage[english]{babel}
\usepackage[utf8x]{inputenc}
\usepackage[T1]{fontenc}
\usepackage{ae,aecompl}
\usepackage{soul}
\usepackage[usenames,dvipsnames]{xcolor}
\usepackage{color}
\definecolor{darkgreen}{rgb}{0,0.6,0}
\definecolor{darkblue}{rgb}{0,0,0.6}
\definecolor{darkred}{rgb}{0.6,0,0}
\definecolor{darkpurple}{rgb}{0.5,0,0.5}

\usepackage{hyperref}

\hypersetup{
	bookmarksopen=true,
	pdftitle=Caballero et al - Microscopic interplay of temperature and disorder of a 1D elastic interface,
	pdfauthor=,
	pdftoolbar=false,           
	pdfstartview={FitH},		
	pdfmenubar=true,			
	pdfhighlight=/O,			
	colorlinks=true,			
	urlcolor=darkblue,
	citecolor=darkblue,		
	linkcolor=darkpurple	
}
\usepackage[normalem]{ulem}

\newcommand{\KPZ}{\text{KPZ}}
\newcommand{\dis}{\text{dis}}
\newcommand{\theo}{\text{th}}
\newcommand{\ther}{\text{th}}

\newcommand{\argp}[1]{\left(#1\right)}
\newcommand{\valabs}[1]{\vert #1\vert}
\newcommand{\moy}[1]{\left\langle  #1 \right\rangle }

\def\de{\mathrm d}

\begin{document}
	
\preprint{APS/123-QED}

\title{Microscopic interplay of temperature and disorder of a one-dimensional elastic interface}
	
\author{Nirvana Caballero}
\email[Corresponding author: ]{Nirvana.Caballero@unige.ch}
\affiliation{Department of Quantum Matter Physics, University of Geneva, 24 Quai Ernest-Ansermet, CH-1211 Geneva, Switzerland}

\author{Thierry Giamarchi}
\affiliation{Department of Quantum Matter Physics, University of Geneva, 24 Quai Ernest-Ansermet, CH-1211 Geneva, Switzerland}

\author{Vivien Lecomte}
\affiliation{Université Grenoble Alpes, CNRS, LIPhy, FR-38000 Grenoble, France}

\author{Elisabeth Agoritsas}
\affiliation{Institute of Physics, Ecole Polytechnique Fédérale de Lausanne (EPFL), CH-1015 Lausanne, Switzerland}


\begin{abstract}
Elastic interfaces display scale-invariant geometrical fluctuations at sufficiently large lengthscales. Their asymptotic static roughness then follows a power-law behavior, whose associated exponent provides a robust signature of the universality class to which they belong. The associated prefactor has instead a non-universal amplitude fixed by
the microscopic interplay between thermal fluctuations and disorder, usually hidden below experimental resolution. Here we compute numerically the roughness of a one-dimensional elastic interface subject to both thermal fluctuations and a quenched disorder with a finite correlation length. We evidence the existence of a novel power-law regime at short lengthscales. We determine the corresponding exponent $\zeta_\dis$ and find compelling numerical evidence that, contrarily to available
analytic predictions, one has  $\zeta_\dis < 1$. We discuss the consequences on the temperature dependence of the roughness and the connection with the asymptotic random-manifold regime at large lengthscales. We also discuss the implications of our findings for other systems such as the Kardar--Parisi--Zhang equation and the Burgers turbulence.


\end{abstract}

\maketitle

\section{Introduction}
\label{sec-introduction}

Interfaces are ubiquitous in Nature and provide remarkable challenges both from theoretical and experimental fronts~\cite{Fisher_review_collective_transport}.
Experimentally they span very different underlying physics and characteristic scales, with examples ranging from domain walls in ferromagnetic or ferroelectric thin films~\cite{lemerle_domainwall_creep,ferre_2013_ComptesRendusPhys14_651,paruch_2013_ComptesRendusPhys14_667,durin_2016_PRL_magneticavalanches,caballero2017excess,pardo2017,salje_2019_PRM_ferroelectricavalanches,tuckmantel_2021_local}
to imbibition fronts in porous media~\cite{alava_2004_AdvPhys53_83},
fracture surfaces~\cite{santucci_2007_PhysRevE75_016104,laurson_NatComm__4_2927_2013cracks,santucci_2019_PhilTransRoyalSocA_377_2136_avalanchescrckling}
and growing fronts of cell colonies~\cite{chepizhko_2016_PNAS113_11408,alert_2020_AnnRev_collectvecellmigration,rapin_2021_roughness}.
On the theory side, describing the competition between the elastic interactions that tend to order them, and the temperature and system heterogeneity that hinder this tendency, is a considerable theoretical challenge, with a resulting out-of-equilibrium physics akin to the one of glasses~\cite{berthier_biroli_2011_RevModPhys83_587}.

A successful theoretical tool to study interfaces is provided by the \emph{disordered elastic systems} framework \cite{fisher_depinning_meanfield,blatter_vortex_review,chauve_creep_short,chauve_2000_ThesePC_PhysRevB62_6241,agoritsas_2012_ECRYS2011,wiese_2021_Arxiv-2102.01215}:
interfaces are modeled as elastic objects evolving in a quenched disordered landscape and subject to thermal noise.
Remarkably, this minimal description is enough to account for many key statistical features of the geometry and dynamics of static or driven interfaces~\cite{ferrero_2021_AnnualReviewsCondMattPhys_creep}. In particular, geometrical fluctuations are scale-invariant at sufficiently large lengthscales, evidenced by a power-law behavior of the roughness, defined as the variance of the relative displacement between two points in the interface.
The associated roughness exponent provides a robust signature of the universality class to which an interface belongs~\cite{Barabasi-Stanley}, depending solely on its dimensionality and on the nature of its elasticity and underlying disorder~\cite{agoritsas_2012_ECRYS2011}.

Beyond this hallmark of universality, the roughness prefactor itself encodes quantitative information about the microphysics of a given system. Labelled as non-universal, this roughness feature has been poorly exploited up to now, despite of its crucial experimental relevance for the quantitative determination of characteristic scales and the validity range of theoretical predictions in a given experimental setup~\cite{jordan_2020_PhysRevB101_184431,cortes_2021_PhysRevB104_144202,rapin_2021_APL_dynamicresponse}.
At equilibrium its amplitude is fixed at short lengthscales by the microscopic interplay between thermal fluctuations and a spatially-correlated disorder.
The resulting temperature crossover below a characteristic energy scale ${T_c(\xi)}$, associated to the disorder strength and finite correlation length $\xi$, is thus a macroscopic `smoking gun' of the spatial structure of microscopic disorder \cite{agoritsas_2010_PhysRevB_82_184207,agoritsas_2012_FHHtri-analytics,agoritsas_2012_FHHtri-numerics}.
Furthermore, studies using perturbative~\cite{korshunov_2013_JETP117_570} or
variational~\cite{agoritsas_2010_PhysRevB_82_184207} methods suggest at short-lengthscales a power-law excess roughness --in addition to thermal fluctuations-- with a characteristic exponent ${\zeta_{\text{dis}} = 1}$.
Unfortunately a quantitative characterization of this microscopic interplay, usually hidden below experimental resolution, has been difficult to access up to now.

For static one-dimensional (1D) interfaces, understanding this interplay is important
to characterize the finite-time and steady-state fluctuations of the 1D
Kardar--Parisi--Zhang (KPZ) equation~\cite{kardar_1986_originalKPZ_PhysRevLett56_889,huse_henley_fisher_1985_PhysRevLett55_2924}.
Considerable advances were achieved recently~\cite{corwin_2011_arXiv:1106.1596,calabrese_exact_2011,doussal_kpz_2012,halpin-healy_takeuchi_2015_JStatPhys160_794,quastel_spohn_2015_JStatPhys160_965}
allowing for the computation of universal exponents and  distributions~\cite{halpin-healy_takeuchi_2015_JStatPhys160_794,quastel_spohn_2015_JStatPhys160_965} for spatially-uncorrelated noises.
Nevertheless, a regime which resists an exact analytical treatment is the low-temperature limit in a spatially-correlated
disorder~\cite{bouchaud_scaling_1995,korshunov_fluctuations_1998,agoritsas-2012-FHHpenta,agoritsas_2012_FHHtri-analytics,agoritsas_2012_FHHtri-numerics,dotsenko_2016_JStatMech2016_123304,agoritsas_lecomte_2017_JPhysA50_104001,mathey_agoritsas_2017_PhysRevE95_032117,dotsenko_2018_JStatMech2018_083302},
despite its relevance to analyze experimental realizations of KPZ~\cite{takeuchi_growing_2011,takeuchi_evidence_2012,halpin-healy_takeuchi_2015_JStatPhys160_794,takeuchi_appetizer_2018},
or of Burgers turbulence~\cite{gotoh_steady-state_1998,bec_burgers_2007}.

In this paper, we address these issues by numerically computing the roughness of a 1D interface with short-range correlated disorder and thermal fluctuations. We unveil the key regime where the interplay between temperature and disorder leads to a roughness excess -- compared to bare thermal fluctuations -- and show that it relaxes towards a power-law behavior ${B_\dis(r) \approx A_\dis (T) \, r^{2 \zeta_\dis}}$ at short lengthscale $r$.
For the specific model we consider, previous analytical but approximate predictions
propose $\zeta_\dis = 1$; yet, we find compelling numerical evidence showing that $\zeta_\dis < 1$,  with important consequences on the temperature dependence of the roughness and the connection to the asymptotic random-manifold regime.

The outline of the paper is the following. After describing the model in Sec.~\ref{sec-model}, we focus on the roughness and disentangle its different contributions in Sec.~\ref{sec-roughness}. We then present a scaling description for the crossovers between different roughness regimes in Sec.~\ref{sec-crossovers-scalings}. We further discuss the implications of our results for KPZ-related systems in Sec.~\ref{sec-implications-others}, and conclude in Sec.~\ref{sec-conclusion}.

\begin{figure}[t]
\begin{center}
{\includegraphics[width=1\linewidth]{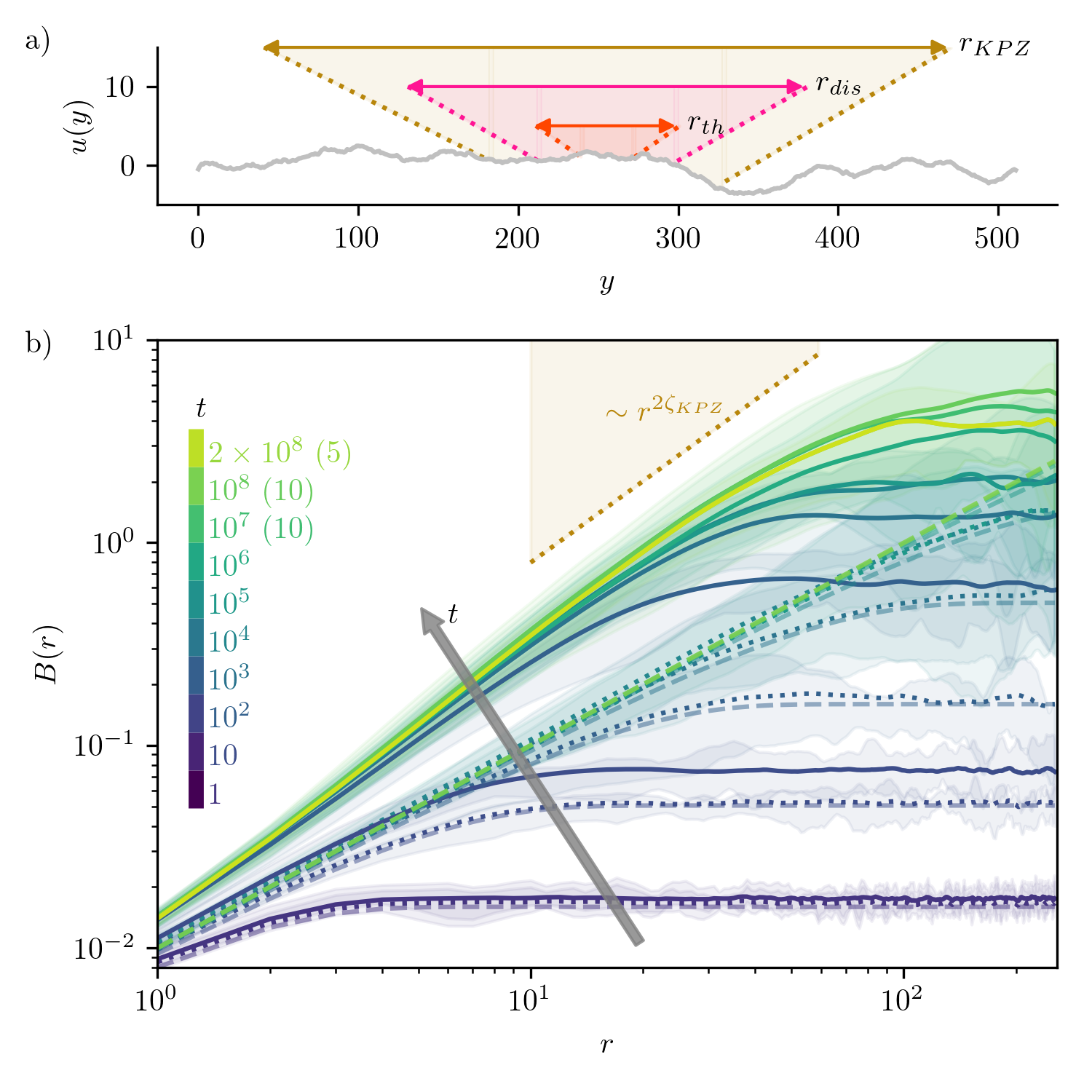}}
\end{center}
\caption{
\label{fig:u_evolution_Br}
a)~Snapshot of an initially flat interface 
$u(y,t)$ evolved under the qEW dynamics~\eqref{eq:qEW} up to time $t=10^9$.
b)~Roughness ${B(r,t)}$ obtained for interfaces that evolved during different times averaged over 50 realizations (10 or 5 realizations for the larger studied times --~indicated in the color scale) with and without disorder (continuous and dotted lines, respectively) at $T=0.01$ and $\epsilon=0.1$.
Shaded fluctuations are shown in the same color scale.
Dashed lines correspond to the analytical prediction (\ref{eq:B(r,t)fromFlat})
in a clean system.
The expected power-law at large distances in the
disordered case (with $\zeta_\KPZ=2/3$) is indicated in a brown dotted line.
}
\end{figure}

\section{Model}
\label{sec-model}

We consider a 1D interface parametrized at time $t$
by ${(y,u(y,t))\in [0,L] \times \mathbb{R} \subset \mathbb{R}^2}$, as shown in Fig.~\ref{fig:u_evolution_Br} a).
The displacement field ${u(y,t)}$ is univalued with periodic boundary conditions ${u(y=0,t)=u(y=L,t)}$.
Starting from a flat initial condition ${u(y,t=0)=0}$ the interface evolves according to the quenched Edwards--Wilkinson (qEW) equation \cite{edwards_wilkinson,kolton_2005_PhysRevLett94_047002,ferrero_2013_ComptesRendusPhys14_641,Ferrero2013nonsteady} in absence of an external force:
\begin{equation}\label{eq:qEW}
 \eta \partial_t u(y,t)= c \partial^2_y u(y,t) + F_p \argp{y,u(y,t)} + \chi(y,t)
 \, .
\end{equation}
Thermal fluctuations are described by a centered Gaussian white noise ${\chi(y,t)}$ of two-point correlator
${\moy{\chi(y_1,t_1) \chi(y_2,t_2)} = 2\eta T \delta(y_2-y_1) \delta(t_2-t_1)}$,
where ${\moy{\dots}}$ denotes the thermal average, $\eta$ is the
friction coefficient and $T$ the temperature (the Boltzmann constant is fixed to ${k_B=1}$).
We fix the units of time and energy by setting $\eta=1$ and the elastic constant $c=1$.
The  disorder is a quenched random potential ${V_p(y,u)}$ and its associated pinning force ${F_p(y,u)=-\partial_u V_p(y,u)}$,
both Gaussian with zero mean and correlators:
\begin{equation}\label{eq:pinningcorrelations}
\begin{split}
 & \overline{V_p(y_1,u_1)V_p(y_2,u_2)}
 	= D \, R_{\xi}(u_2-u_1) \, \delta(y_2-y_1)
 \, , \\
 & \overline{F_p(y_1,u_1)F_p(y_2,u_2)}
	= \Delta_{\xi} (u_2-u_1) \, \delta(y_2-y_1)
 \, ,
\end{split}
\end{equation}
where $\overline{\vphantom{|}\cdots\vphantom{|}}$ denotes the average over disorder realizations.
We consider the case of `random-bond' disorder with finite correlation length~${\xi}$, \textit{i.e.}~with a short-range correlator ${R_\xi(u)=\xi^{-1} R_1(u/\xi)}$ normalized as ${\int_{\mathbb{R}} \de u \, R_{\xi}(u)=1}$ and $D$ the disorder strength. Both correlators are even and related through ${\Delta_{\xi}(u) = - D \, R_{\xi}''(u)}$ \cite{chauve_2000_ThesePC_PhysRevB62_6241}, so that ${\int_{\mathbb{R}}\de u \, \Delta_{\xi}(u)=0}$.

We numerically integrate (\ref{eq:qEW}) keeping $u$ as a continuous variable while discretizing the $y$ direction in $L=512$ segments of unit length.
These qEW settings allow us to advantageously replace the Dirac ${\delta(y-y')}$ by the Kronecker ${\delta_{y y'}}$ in (\ref{eq:pinningcorrelations}) and thus to implement an uncorrelated disorder along the internal direction $y$.
In the transverse direction, the disorder potential is dynamically generated \cite{salmon2011parallel,Ferrero2013nonsteady} with random numbers taken from a uniform distribution in the range $[-\frac{\epsilon}{2},\frac{\epsilon}{2}]$ at equidistant positions, spaced by ${\Delta u=1}$.
To obtain the random-bond pinning force at a point $u$, we interpolate the two nearest random numbers with a linear spline and take its derivative with respect to $u$.
With these settings, the correlator ${\Delta_\xi(u=u_2-u_1) = \frac23 \frac{D}{\xi^3} \Delta_{\text{adim}} (\hat{u}=u/\xi)}$ is fully given by a piecewise linear continuous function, ${\Delta_{\text{adim}} (\hat{u})}$, that connects the values
\begin{equation}
\left\lbrace \begin{array}{l}
\Delta_{\text{adim}} (\hat{u}=0)=2
\\
\Delta_{\text{adim}} (\vert \hat{u} \vert=1)=-1
\\
\Delta_{\text{adim}} (\vert \hat{u} \vert \geq 2)=0
\end{array} \right.
\end{equation}
(see Appendix~\ref{appendix-pinning-force-correlator-disorder-strength}). 
Our effective parameters are then $\lbrace \xi=1, D=\frac{\epsilon^2}{12} \rbrace$, where we took $\epsilon=0.1$,
and we explore the temperature range ${T \in \lbrace 0.005, \dots, 0.074 \rbrace}$.
Further technical details on the numerical integration are reported in Appendix~\ref{appendix-parameters-numerical-simulations}.

We emphasize that we explicitly consider a spatially-correlated disorder in the transverse direction, \textit{i.e.}~allowing for a resolution at displacements smaller than ${\xi}$. This is a key difference compared to other numerical studies of the qEW dynamics, which usually assume a discrete description of the interface on a lattice, with a disorder completely uncorrelated from site to site.
The latter allows for highly efficient computations to access roughness scalings at asymptotically large lengthscale~\cite{kardar_1985_PhysRevLett55_2923,kardar_1987_PhysRevLett58_2087,ferrero_2013_ComptesRendusPhys14_641}, but by definition it prevents any study of the microscopic interplay between thermal fluctuations and disorder. For equilibrated 1D interfaces, an alternative approach is provided by the `1+1 directed polymer' in a correlated disorder with ${\xi>0}$, studied through the KPZ equation that its free energy satisfies \cite{agoritsas_2012_FHHtri-numerics}. The latter requires, however, the disorder to have a finite correlation length along the internal direction $z$ and bears by construction an artifact at short lengthscales --these two aspects becoming increasingly problematic at lower temperatures.
Our numerical approach bypasses these issues in order to provide a reliable characterization, for the qEW dynamics~\eqref{eq:qEW}, of the temperature-dependent microscopic interplay we are interested in.

\section{Disentangling roughness contributions} 
\label{sec-roughness}

To characterize the geometrical fluctuations of the interface we focus on the
roughness function (also called `height-height correlation function' in the context of growing surfaces \cite{Barabasi-Stanley}):
\begin{equation}
\label{eq-roughness-def}
B\argp{r=\valabs{y_2-y_1},t} = \overline{ \moy{ [u(y_2,t)-u(y_1,t)]^2}} \, .
\end{equation}
It quantifies the variance of the relative displacements of the interface, as a function of the lengthscale~$r$, and inherits the translation invariance in $y$ of the microscopic disorder.
For a clean system
($F_p(y)=0$) we can compute analytically the time dependence of this correlation for an infinite interface \cite{edwards_wilkinson,caballero_2020_PhysRevB102_104204}:
\begin{equation}\label{eq:B(r,t)fromFlat}
 B_\ther(r,t)
 =\frac{Tr}{c} \bigg[1- \frac{1}{\sqrt{\pi}zr}\Big(e^{-z^2r^2}-1\Big)- \frac{2}{\sqrt{\pi}}\int_0^{zr}
   \!\!\!\!\!\!
\de t\,e^{-t^2} \bigg]
\end{equation}
where ${z=\sqrt{\frac{\eta}{8ct}}}$.
At large times, (\ref{eq:B(r,t)fromFlat}) converges to the static thermal roughness ${B_\ther(r) \equiv \frac{Tr}{c}}$.
To disentangle different contributions on $B\argp{r,t}$, we introduce the excess roughness ${B_{\dis} = B- B_{\ther}}$, defined as the difference between the total roughness and its analytical value in the clean case.
Note that for a finite size $L$ with periodic boundary conditions, we have ${B_\ther (r,L) = \frac{Tr}{c}(1-r/L)}$, thus predicting a saturation for lengthscales ${r \lesssim L/2}$ and a decrease of the roughness beyond (also expected for the whole roughness function).

In Fig.~\ref{fig:u_evolution_Br}, we show the  evolution of $B(r,t)$ at a fixed temperature in clean and disordered systems. For the clean case, results match the analytical prediction (\ref{eq:B(r,t)fromFlat}).
For the disordered case, at very large times and large lengthscales, the power-law ${\sim r^{2\zeta_\KPZ}}$ with $\zeta_\KPZ=\frac{2}{3}$,  characteristic of a random-bond disorder is reached.
Note, however, that already after $t=10^3$ the roughness has converged at short lengthscales ($r\lesssim 20$).

\begin{figure}
\begin{center}
\includegraphics[width=1\linewidth]{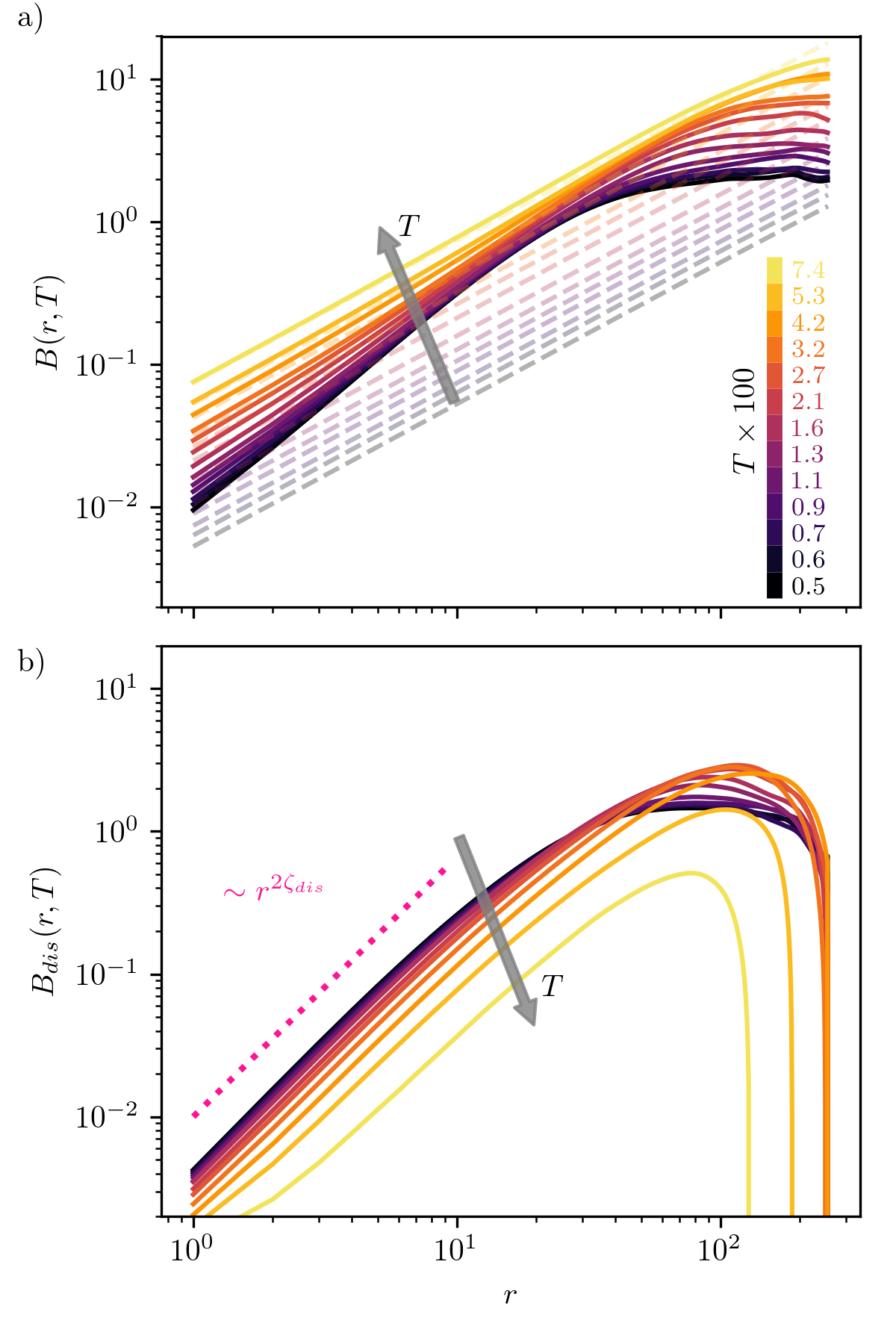}
\end{center}
\caption{\label{fig:Br_Bdisr_T}
a)~Roughness of interfaces evolving in a spatially-correlated disorder,
for different temperatures $T$.
Each curve corresponds to an average over realizations that evolved with the qEW dynamics \eqref{eq:qEW} for a time $t=10^6$ starting from a flat initial condition.
Dashed lines represent the roughness (\ref{eq:B(r,t)fromFlat}) in absence of disorder.
 b)~Corresponding excess roughness $B_\dis(r)$ due to disorder, obtained as the difference between the total roughness and the thermal roughness (\ref{eq:B(r,t)fromFlat}). The decrease beyond ${r=L/2}$ is an artifact of the periodic boundary conditions. Our numerical study shows the existence of a power-law regime for the excess roughness at small scales, characterized by an exponent $\zeta_\dis$.}
\end{figure}

In Fig.~\ref{fig:Br_Bdisr_T} a), we show $B(r,T)$ at fixed large time $t=10^6$ averaged over 50 realizations \cite{bustingorry_agoritsas_2021_JPhysCondensMatter33_345001}. To increase the statistics, for each realization that evolved for a time $10^6$, we included 100 more configurations equally spaced in time intervals of $10^3$. This procedure allows us to increase the statistics for lengthscales at which the interface has already equilibrated.
In Fig.~\ref{fig:Br_Bdisr_T} b), we report the corresponding excess roughness ${B_\dis(r,T)}$ at different temperatures. The numerical data shows the existence of a power-law regime of ${B_\dis(r,T)}$ characterized by an exponent $\zeta_\dis$. This regime, which can be obscured by the existence of the large thermal component of the roughness, is nevertheless present and universal~\footnote{%
In Appendix~\ref{appendix-disorder-Gaussian-distribution}, we compare our results for two very different disorder distribution, generated from potentials which have bounded and unbounded support.
We show that ${B_\dis(r,T)}$ presents the same exponent $\zeta_\dis$, providing compelling evidence for the universality of the power-law regime characterized by this exponent.
}, and results from the interplay between the finite correlation of the disorder and the thermal fluctuations.
As we discuss in Sec.~\ref{sec-crossovers-scalings} and Sec.~\ref{sec-implications-others}, besides the existence of the regime itself, the value of the exponent $\zeta_\dis$ has important consequences for the temperature dependence of the roughness.

\begin{figure}
\begin{center}
\includegraphics[width=1\linewidth]{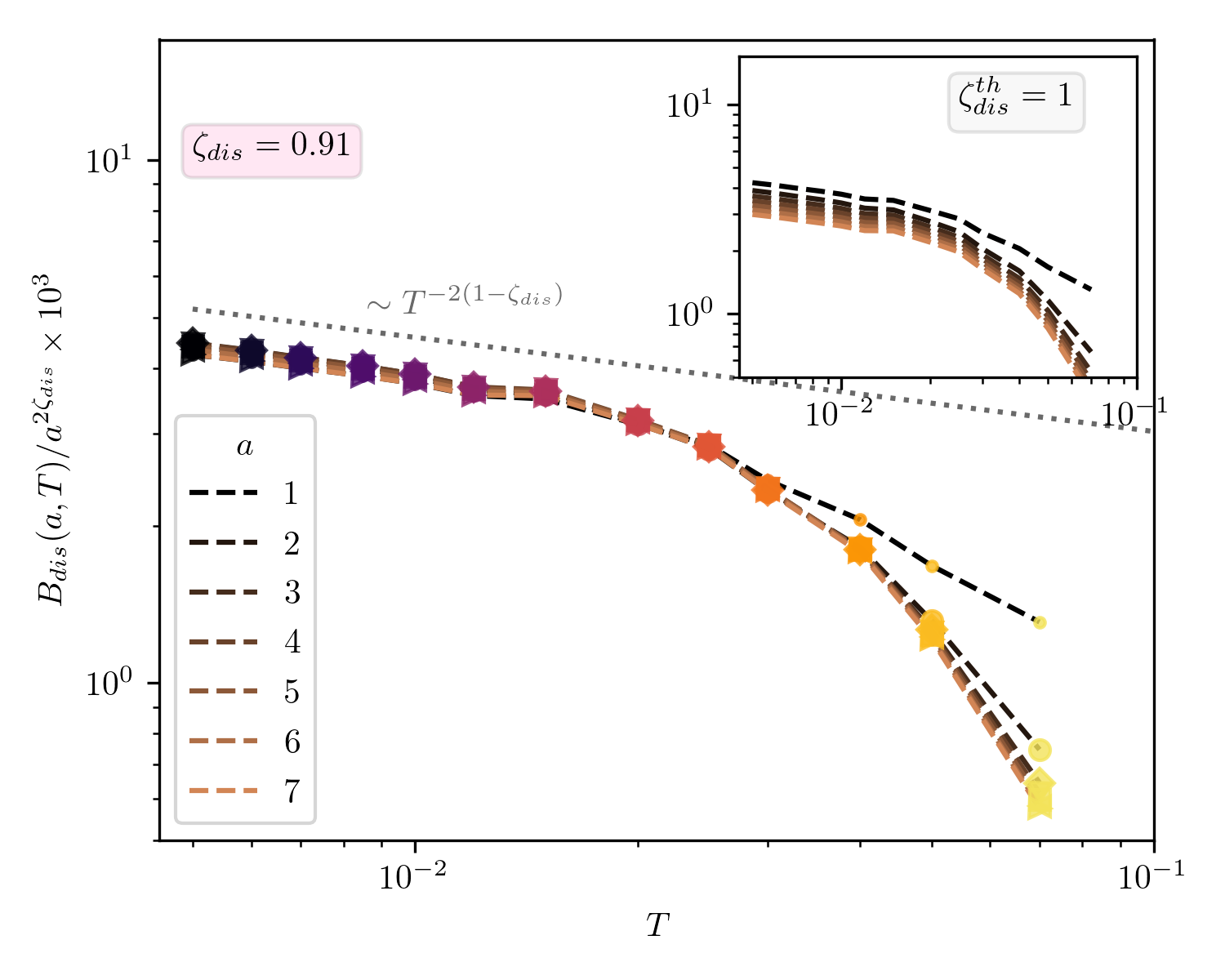}
\end{center}
\caption{\label{fig:scaling}
Collapse of the excess roughness $B_\dis(a,T)$ for different fixed lengthscales $a$ as a function of temperature. The collapse is obtained with a power-law $a^{2\zeta_\dis}$, with $\zeta_\dis=0.91\pm 0.05$. The insert shows that the collapse is lost for ${\zeta_\dis^\theo=1}$.}
\end{figure}

A fit of $B_\dis(r)$ with a power-law ${\sim r^{2\zeta_\dis}}$ in the regime of small values of $r$ gives an exponent $\zeta_\dis=0.91$ with a temperature-dependent amplitude. Given the importance of estimating if $\zeta_\dis=1$, we now explore this issue.
This result is further detailed in Fig.~\ref{fig:scaling} where the rescaling of $B_\dis$ with the power-law $r^{2\zeta_\dis}$ is shown.
The collapse of the curves is very sensitive to the precise value of the exponent and allows us to rule out with a high confidence a value of $\zeta_\dis=1$ (see Appendix~\ref{appendix-determination-best-value-zetadis}).
Interestingly, the only currently available analytical predictions gives ${\zeta_\dis=\zeta_\dis^\theo\equiv 1}$, in distinct methods:
a finite-temperature perturbative approach~\cite{korshunov_2013_JETP117_570} and two different computations based on a Gaussian-Variational-Method (GVM)~\cite{agoritsas_2010_PhysRevB_82_184207}.
Our numerical findings thus provide evidence that non-perturbative approaches, beyond the usual variational alternatives, are required to determine $\zeta_\dis$. We also checked that a different disorder distribution leads to the same value of $\zeta_\dis$ (see Appendix~\ref{appendix-disorder-Gaussian-distribution}), supporting the robustness of our finding that ${\zeta_\dis<1}$.

\section{Crossovers between power-law roughness regimes}
\label{sec-crossovers-scalings}

To analyze the implications of this result, we determine the relevant lengthscales.
The static roughness ${B(r)=B_\ther(r)+B_\dis(r)}$
crosses over from a thermal regime ${B(r) \approx B_\ther(r)=T r^{2 \zeta_\ther}/c}$ at very short lengthscales to the random-manifold regime  ${B(r) \approx B_\dis(r) \approx \xi_{\text{eff}}^2 \, (r/L_c)^{2 \zeta_\KPZ}}$ at lengthscales larger than a characteristic length $L_c$,
closely related to the `Larkin length' in higher-dimensional interfaces \cite{Larkin_model_1970-SovPhysJETP31_784,blatter_vortex_review}.
In our system, ${\zeta_\ther=\frac12}$ and ${\zeta_\KPZ=\frac23}$.
Previous analytical studies
\cite{agoritsas_2010_PhysRevB_82_184207,agoritsas_2012_FHHtri-analytics,agoritsas_lecomte_2017_JPhysA50_104001}
have shown that
${L_c=\frac{(T/f)^5}{cD^2}}$ and ${\xi_{\text{eff}}=\frac{(T/f)^3}{cD}}$,
with a temperature-dependent parameter $f$.
Defining the energy scale ${T_c=(\xi c D)^{1/3}}$
we expect a monotonous crossover from ${f \approx 1}$ at ${T \gg T_c}$
to ${f \sim T/T_c}$ at ${T \ll T_c}$ \footnote{See Eqs.~(49), (97) in \cite{agoritsas_2012_FHHtri-analytics}}.
The \emph{effective} characteristic energy scale ${\widetilde{E}=T/f}$ controls all the relevant scales for the geometrical fluctuations and in the two previous limits, is fixed by thermal fluctuations (${\widetilde{E} \approx T}$) or by disorder (${\widetilde{E} \approx T_c}$).
From our numerical findings ${B_\dis(r)}$ has two power-law regimes with a crossover lengthscale $r_0$:
\begin{equation}\label{eq-Bdis-2powerlaws-crossover}
 B_\dis(r) = \left\lbrace \begin{array}{ll}
 	A_1 (r/r_0)^{2 \zeta_\dis} : & r \ll r_0 \, ,
 	\\
 	A_2 (r/r_0)^{4/3} : & r \gg r_0 \, .
 	\end{array} \right.
\end{equation}
Let us fix $r_0$ with a scaling argument to determine  ${\lbrace L_c,\xi_{\text{eff}} \rbrace}${~\cite{agoritsas_lecomte_2017_JPhysA50_104001}}.
We rescale the spatial coordinates ${(y=b \bar{y},u=a \bar{u})}$ and parameters ${\lbrace c=c',D=D_0 D',T=\widetilde{E} T',\xi=a \xi' \rbrace}$ while leaving the Boltzmann weight invariant so that ${B(r;c,D,T,\xi)=a^2 B(r/b;c',D',T',\xi')}$.
To focus on the microscopic interplay of temperature and disorder, we fix the scales ${a=\xi}$ and ${\widetilde{E}=T/f}$ (\textit{i.e.}~${\lbrace \xi'=1,T'=f \rbrace}$)
which implies ${D_0=\frac{(T/f)^3}{c \xi}}$ and ${b=\frac{c \xi^2}{T/f}}$ \footnote{The high-temperature version of this scaling choice was reported in Table~(2a) of \cite{agoritsas_lecomte_2017_JPhysA50_104001}, but had an unpleasant ill-defined limit at asymptotically \emph{large} lengthscales for both ${T \to 0}$ and ${\xi \to 0}$. We argue here that it is instead relevant for \emph{short} lengthscales.}.
The short-lengthscale regime where the disorder correlation length and the effective energy scale $\widetilde{E}$ are equally relevant lies at ${r \leq b}$, so we can identify ${r_0=b}$.
Note that the effective disorder strength ${D'=\frac{T_c}{T} f}$ crosses over from ${D' \approx 1}$ at ${T \ll T_c}$ to ${D' \sim T_c/T}$ at ${T \gg T_c}$. As physically expected, the latter case supports a small-disorder perturbative expansion at  high temperature.

We thus can determine the prefactors for ${B_\dis(r \leq r_0)}$ in (\ref{eq-Bdis-2powerlaws-crossover}):
${A_1=B_\dis(r_0)=A_2}$ with $A_2$  fixed from the large scales from ${A_2/r_0^{4/3}=\xi_{\text{eff}}^2/L_c^{4/3}}$.
Using ${r_0=\frac{c \xi^2}{T/f}}$ we get
\begin{equation}\label{eq:Adis}
 A_\dis(T)
 = \frac{B_\dis(r \leqslant r_0)}{r^{2 \zeta_\dis}}
 \propto \frac{\xi^2 T_c^2}{\argp{c \xi^2}^{2 \zeta_\dis}} \argp{\frac{f}{T}}^{2(1- \zeta_\dis)}
\end{equation}
Note that this implies that ${\zeta_\dis=\zeta_\dis^\theo=1}$ leads to a temperature-independent roughness prefactor.
On the contrary, a value ${\zeta_\dis \neq 1}$, as we find numerically, implies that ${A_\dis(T)}$ \emph{does} depend on temperature.
We report this feature in Fig.~\ref{fig:scaling}, thus further supporting $\zeta_\dis<1$.

From the behavior of the interpolating parameter
${f \approx 1}$ at ${T \gg T_c}$
and
${f \sim T/T_c}$ at ${T \ll T_c}$,
Eq.~\eqref{eq:Adis} implies that $A_\dis(T)$
decreases with increasing $T$ at ${T \gg T_c}$ and saturates at ${T \to 0}$.
This scenario is at variance with our numerical findings in Fig.~\ref{fig:scaling}: $A_\dis(T) $ does indeed present a regime of temperature $\sim T^{-2(1-\zeta_\dis)}$ compatible with (\ref{eq:Adis}).
However, such power-law behavior is expected to hold at high $T$ and to provide an upper bound to the low-$T$ saturation. The reasons for this discrepancy are unclear  but could stem from:
\textit{(i)}~numerical prefactors in the estimation of $T_c(\xi)$,
so that the regime ${T\ll T_c}$ would in fact be reached below our temperature range;
\textit{(ii)}~the discretization along the internal direction $y$,
which could add one lengthscale to take into account in the scaling analysis. A breakdown of the above scaling argument cannot be excluded, but ${r_0={c\xi^2}f/{T}}$ does coincide with a more involved prediction obtained in the KPZ language \footnote{See Eq.~(99) in \cite{agoritsas_2012_FHHtri-analytics}, where the DP `time' ${t_{\text{sat}}}$ can be identified with our crossover $r_0$.}.
This discrepancy, which does not question the $\zeta_\dis$ roughness regime existence requires further investigations that are beyond our present scope.

Let us now turn to the case of very small thermal fluctuations at $T \to 0^+$, where the crossover between the thermal and $\zeta_{\text{dis}}$ power-law regimes can be further examined analytically and numerically.
The crossover lengthscale $r_0$ is always bounded by the Larkin length $L_c$:
\begin{equation}
\left\lbrace \begin{array}{ll}
 T \gg T_c : & r_0 \ll L_c \, , \text{ with } r_0 \sim \frac{c \xi^2}{T} \, , \, L_c \sim \frac{T^5}{cD^2} \, ,
 \\
 T \ll T_c : & r_0 \lesssim L_c \sim \frac{T_c^5}{cD^2} = \frac{c^{2/3} \xi^{5/3}}{D^{1/3}} \, .
\end{array} \right.
\end{equation}
At high temperature, the regime ${r \lesssim r_0}$ where ${B_\dis(r)\sim r^{2\zeta_\dis}}$ has a small extension and is screened by the thermal contribution $Tr/c$ in $B(r)$; hence, as $r$ increases, the total roughness $B(r)$  crosses over directly from the thermal to the random-manifold regime, with a single crossover at the high-$T$ Larkin length $L_c$ such that ${B_\ther(L_c) \approx B_\dis(L_c)}$. 
On the contrary, in the limit ${T \to 0^+}$ we have ${B(r) \approx B_\dis(r)}$, which crosses over from the new power-law regime that we characterized ($\zeta_\dis$) to the well-studied random-manifold regime (${\zeta_\KPZ}$), and a vanishing thermal regime at ${r \leq r_1 \sim T^{1/[1-2(1-\zeta_\dis)]}}$ \footnote{At low $T$, ${B_\ther(r)}$ intersects ${B_\dis(r)}$ at a crossover ${r_1<r_0}$.
We can evaluate $r_1$ at ${T \ll T_c}$ from ${T r_1/c \equiv A_1 (r_1/r_0)^{2 \zeta_\dis}}$, which yields
${r_1^{1-2(1-\zeta_\dis)} = \frac{T \, (T/f)^{4/3}}{c \xi^{4/3} D^{2/3}} r_0^{2(1-\zeta_\dis)}}$ with ${T/f \sim T_c}$}.
By studying the lengthscale $r_1$ (see Appendix~\ref{appendix-scaling-crossover}), we find our numerical results to be compatible with a scaling $\zeta_\dis\approx 0.91$, but not with $\zeta_\dis=\zeta_\dis^\theo=1$.

\begin{figure}[!h]
    \centering
    \includegraphics[width=1\linewidth]{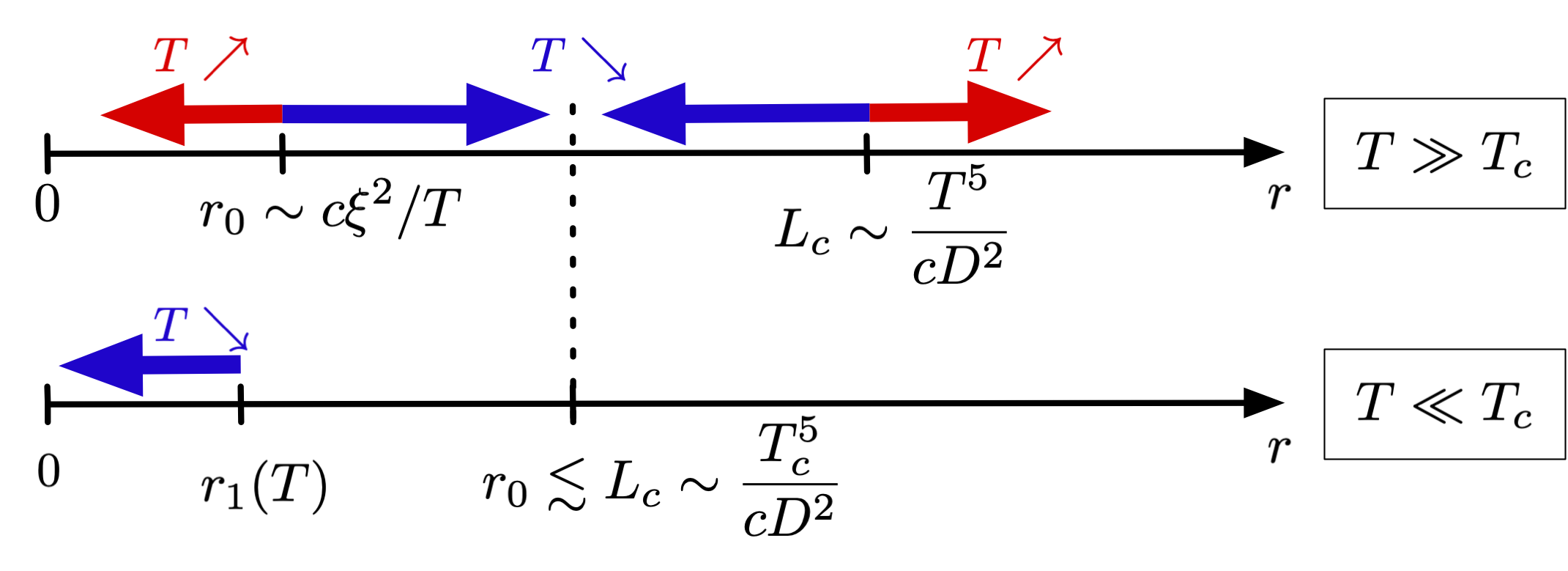}
    \caption{Summary of the crossover scales at high \textit{versus} low temperature for the equilibrated 1D qEW interface.
    Red (respectively blue) arrows indicate how crossover scales change when the temperature increases (respectively decreases), in both temperature regimes.
    We always have ${r_0 \lesssim L_c}$.
    Note that assuming ${\xi=0}$ corresponds to setting ${T_c=0}$, in which case we could account only for the high-temperature regime.}
    \label{fig:summary-crossovers}
\end{figure}

Physically, the characteristic lengthscales ${\lbrace L_c , r_0, r_1 \rbrace}$ and the energy scale ${\widetilde{E}=T/f}$ are related to the microscopic interplay between temperature and disorder. The Larkin length can be deduced from asymptotically large lengthscales, easier to access analytically and experimentally, hence its crucial role in previous studies \cite{Larkin_model_1970-SovPhysJETP31_784,blatter_vortex_review,chauve_2000_ThesePC_PhysRevB62_6241,agoritsas_2010_PhysRevB_82_184207,agoritsas_2012_FHHtri-analytics,agoritsas_lecomte_2017_JPhysA50_104001}.
In this paper, we argue instead that it stems as a consequence of short-lengthscale properties, usually hidden below experimental resolution, that one can nevertheless investigate via the excess roughness due to disorder.
We expect this scenario --relating the microscopic interplay to the quantitative amplitude of the macroscopic fluctuations-- to be valid beyond the specific interface (1D qEW) we considered.

\section{Implications for other KPZ-related problems}
\label{sec-implications-others}

The relevance of these results extends well beyond the sole equilibrium 1D interface, thanks to the exact mapping onto the so-called 1+1 directed polymer (DP) and the KPZ settings \cite{kardar_1986_originalKPZ_PhysRevLett56_889,huse1985pinning}.
An interface segment can be seen as a polymer growing along a DP `time': fixing an extremity of the DP at the origin (\textit{i.e.} ${u(0)=0}$), its endpoint distribution encodes, at a fixed `time' $r$, the geometrical fluctuations of an equilibrated interface at a lengthscale $r$ and \emph{in a given disorder realization}. The DP free-energy, \textit{i.e.}~the logarithm of this distribution, then obeys a KPZ equation with a `time'~$r$ and a spatially-correlated noise~$V_p$ (our random potential), with a `sharp-wedge' initial condition \cite{huse1985pinning}. In this language, the asymptotic random-manifold scaling at ${r \to \infty}$ describes the approach to the KPZ steady-state regime, and the microscopic interplay at short lengthscales corresponds to the KPZ evolution at short `times'.

The DP formulation allows one to disentangle the thermal and sample-to-sample fluctuations (thermal and disorder averages, respectively). It provides in particular a natural decomposition of the roughness as ${B(r)=B_{\text{th}}(r) + B_{\text{dis}}(r)}$, with a direct definition of the excess roughness as
the second \emph{disorder} cumulant ${B_{\text{dis}}(r)=\overline{\langle u(r)-u(0) \rangle^2}^c}$ \footnote{See Eqs.~(B.12)-(B.13) in \cite{agoritsas_2012_FHHtri-analytics}, derived from the statistical tilt symmetry. Note that this relation is valid under the condition that ${u(0)=0}$ for any given disorder realization $V_p$; since our numerical approach does not impose such a boundary condition, we cannot directly evaluate $B_\dis(r)$ from such an identity.}.
%
%
Such writing in fact encodes a more general decomposition: in a given disorder realization, the DP free-energy can be decomposed as the sum of a purely thermal contribution (that remembers the initial condition $u(0)=0$) and a disorder contribution (that is invariant by translation in distribution and thus forgets --statistically-- the initial condition).
Such a manifestation of the so-called `statistical tilt symmetry' (STS) (valid at $\xi=0$ and $\xi>0$~\footnote{See Appendix B in Ref.~\cite{agoritsas_2012_FHHtri-analytics}.}) means that $B_\dis(r)$ directly encodes the two-point correlations of the DP endpoint when the thermal fluctuations (that fully capture the initial condition) have been subtracted. This explains why $B_\dis(r)$ gives access to a regime of fluctuations that is inherent to disorder and that, in spite of its seemingly artificial definition, it bears a \emph{bona fide} physical content.
We expect that the excess roughness $B_\dis(r)$ can be generalized to other systems presenting STS (\textit{e.g.}~for interfaces in higher dimensions or that are discretized in the longitudinal direction).

The DP free-energy steady-state properties at ${\xi>0}$ have been studied through non-perturbative functional-renormalization-group~\cite{mathey_agoritsas_2017_PhysRevE95_032117},
but the understanding of the short-DP-`time' regime is incomplete, either through variational \footnote{See Eqs.~(56), (126) in \cite{agoritsas_2010_PhysRevB_82_184207}, or Eqs.~(6.21), (6.26) in \cite{phdthesis_Agoritsas2013}.},
perturbative \footnote{See Eq.~(20) in \cite{korshunov_2013_JETP117_570} for the second-order correction at finite temperature, leading to the conjecture stated in Eq.~(21).},
or indirect \footnote{See Sec.~VI in \cite{agoritsas_2012_FHHtri-analytics} and the summary in Fig.~7.1 in \cite{phdthesis_Agoritsas2013}.} approaches.
On the numerical side, we have previously reported in Ref.~\cite{agoritsas_2012_FHHtri-numerics} an apparent power-law behavior for $B_{\text{dis}}$ at short lengthscales, with $\zeta_{\text{dis}} \in [2,2.5]$ depending on the temperature. These exponent values cannot be directly compared to the one found in the present paper for two reasons: 
\textit{(i)}~In Ref.~\cite{agoritsas_2012_FHHtri-numerics}, a singularity at initial KPZ `time', due to the approach used, had to be amended through a short-time regularization that can affect $B_\dis$ in that regime.
\textit{(ii)}~In Ref.~\cite{agoritsas_2012_FHHtri-numerics}, the disorder presented a finite correlation length  $\xi_z$ in the internal direction, with ${\xi_z \approx 1}$ so that ${r_0 \approx \xi_z}$;
this is at odds with the assumption ${\xi_z=0}$ inherent to Eq.~\eqref{eq:pinningcorrelations}, especially because $\xi_z$ is larger than the typical lengthscales of the regime studied in the present paper.
In fact, addressing these issues partly triggered  the present study, whose settings also have the advantage to apply to generic non-stationarized interfaces (and not exclusively to the specific static 1D interface of Ref.~\cite{agoritsas_2012_FHHtri-numerics}).

The scope of the results obtained in the present paper can thus be extended as follows.
Consider the two-point correlator 
of the derivative of the DP free-energy at `time' $r$.
It is a central quantity that encodes the scalings of the amplitude of the KPZ field.
Strictly at $\xi=0$, it contains a singular Dirac delta contribution, that is rounded at finite $\xi>0$:
its properties thus reflect how the disorder correlation length $\xi$ affects the KPZ scaling.
In Ref.~\cite{agoritsas_2012_FHHtri-analytics}, it was shown that scale $r_0$ corresponds to the typical DP `time' at which the maximum of this correlator saturates to its steady-state value.
The short-lengthscale power-law regime of $B_\dis(r)$ governed by $\zeta_\dis$ that we uncovered describes how the DP free-energy (\textit{i.e.}~the KPZ field) distribution evolves at short times.
This indicates that, to access the KPZ scalings at short times, one needs to disentangle the disorder from the thermal fluctuations, and that this happens in the scale-invariant regime governed by $\zeta_\dis$.

Since ${r_0 \to 0}$ as ${T \to 0}$,
the range of this regime vanishes in  high-temperature limit ${T \gg T_c}$  (see Fig.~\ref{fig:summary-crossovers}).
Nevertheless, already in this limit, the KPZ equation at short times with different initial conditions displays distinct large deviations \cite{hartmann_2020_PhysRevE101_012134}:
we thus expect that at finite $T$ and $\xi$, the short-time regime we have uncovered will have a counterpart in such large-deviation analysis.
How these features, and more generally the exact predictions for the KPZ equation available at ${\xi=0}$ \cite{takeuchi_2018_PhysicaA504_77}, are modified for a given disorder correlator ${R_{\xi>0}}(u)$, in the low-temperature limit, remains currently an open issue.
Although the asymptotic roughness exponent ${\zeta_{\text{KPZ}}=\frac23}$ is known to be robust to finite $\xi$ \cite{agoritsas_lecomte_2017_JPhysA50_104001},
the functional form of the KPZ steady-state distribution and correlators --as well as their overall amplitude and characteristic scales-- will be modified by the short-`time' evolution.
For instance, in experiments of growing interfaces in liquid crystals which exhibits geometrical fluctuations remarkably compatible with the KPZ predictions at ${\xi=0}$ \cite{takeuchi_growing_2011,takeuchi_evidence_2012,takeuchi_2018_PhysicaA504_77}.
In particular, short-time fluctuations could be probed within our approach for the droplet initial condition (through the excess roughness or more generally after removing a thermal-like contribution to the KPZ field).
Also, in the large-time regime,
their two-point correlator has a voltage-dependent amplitude which could be a signature of our `low-temperature' regime \footnote{See Sec.~7.2.2 in \cite{phdthesis_Agoritsas2013}.}.

Finally, since the noisy Burgers equation is derived from the KPZ one, our results translate to the Burgers turbulence~\cite{bec_burgers_2007}
where the small-temperature asymptotics corresponds to the inviscid limit and large-scale forcing~\cite{gotoh_steady-state_1998,dotsenko_2018_JStatMech2018_083302}.
In particular, the regime we have identified could be compared to the recent results of Ref.~\cite{cartes_galerkin-truncated_2022},
where, within a Fourier approach, a fluctuation regime scaling as $\sim k^{-1}$ (with $k$ the Fourier mode) was identified numerically
and could interestingly be analysed with other scaling exponents.
%

\section{Conclusion}
\label{sec-conclusion}

In this work,
we find that, hidden under thermal fluctuations of a 1D interface, there exists a scale-invariant regime at short lengthscales governed by an exponent $\zeta_\dis$.
Our numerical analysis shows that ${\zeta_\dis < 1}$, at variance with the existing analytical estimates for such exponent $\zeta_\dis^\theo=1$.
The discrepancy hints towards the non-perturbative nature of the short-lengthscale regime of ${B_\dis(r)}$, which thus requires further analytical investigations.
In systems where thermal fluctuations are strong enough, the scaling regime described by $\zeta_\dis$ is hidden under thermal fluctuations. When the temperature becomes small or when the disorder correlation length becomes large, the power-law behavior $B(r)\sim r^{2\zeta_\dis}$ can be manifested on experimentally or numerically accessible ranges.
For instance, large roughness exponents were found in a recent modelization~\cite{caballero_JSTAT_2021} of experimental measurements~\cite{domenichini2019transient} of magnetic domains subjected to an AC field. A possible cause can be that the AC field effectively increases the disorder correlation length in the sample, thus unveiling the $B(r)\sim r^{2\zeta_\dis}$ regime.
It is worth investigating the existence of such scaling regime in interfaces, and to test its robustness to features such as overhangs (for instance in disordered Ginzburg--Landau models~\cite{caballero2018magnetic}, known to reduce to qEW~\cite{caballero_2020_PhysRevB102_104204}).
Adding a driving force generates a velocity in a non-linear `creep regime', whose scaling is controlled by the static geometrical exponents. It is natural to wonder if the scaling regime we unveiled has consequences for the creep regime, for instance in avalanches statistics.
The case of higher dimensions random manifolds is also open.

\begin{acknowledgments}
We thank {discussions with} Jean-Pierre Eckmann and Sebastian Bustingorry.
This work was supported in part by the Swiss National Science Foundation under Division II.
N.C.~acknowledges support from the Federal Commission for Scholarships for Foreign Students for the Swiss Government Excellence Scholarship (ESKAS No.~2018.0636).
E.A.~acknowledges support from the Swiss National Science Foundation by the SNSF Ambizione Grant PZ00P2{\_}173962.
V.L.~thanks the Université de Genève (where part of this work was performed) for its warm hospitality, and acknowledges support by the ERC Starting Grant No.~680275 MALIG, the ANR-18-CE30-0028-01 Grant LABS and the ANR-15-CE40-0020-03 Grant LSD.

The simulations were performed at the University of Geneva on the \textit{Mafalda} cluster of GPUs.
\end{acknowledgments}

\appendix 

\section{Pinning force correlator and disorder strength}
\label{appendix-pinning-force-correlator-disorder-strength}

In this appendix, we detail how we generate the quenched random potential ${V_p(y,u)}$ and its associated pinning force ${F_p(y,u) \equiv - \partial_u V_p(y,u)}$.
We recall that the interface is parametrized by the univalued displacement field ${u(y,t)}$,
with the internal coordinate $y$ taking discrete values ${y=j \Delta y}$ (where $j$ is an integer, ${j=0, \dots L_y/\Delta y}$),
and the transverse coordinate $u$ taking continuous values.
We consider a disorder with Gaussian distribution fully characterized by a zero mean and the two-point correlators:
\begin{equation}
\label{eq:pinningcorrelations-bis}
\begin{split}
 & \overline{V_p(y_1,u_1)V_p(y_2,u_2)}
 	= D \, R_{\xi}(u_2-u_1) \, \delta_{y_2 y_1}
 \, , \\
 & \overline{F_p(y_1,u_1)F_p(y_2,u_2)}
	= \Delta_{\xi} (u_2-u_1) \, \delta_{y_2 y_1}
 \, ,
\end{split}
\end{equation}
where $\overline{\vphantom{|}\cdots\vphantom{|}}$ denotes the average over disorder realizations.
We chose the normalization ${\int_{\mathbb{R}}\de u \, R_\xi(u)=1}$,
and the correlators are simply related by ${\Delta_\xi(u)=-D R_{\xi}''(u)}$.
In the following, we thus make explicit the functional ${\Delta_\xi(u)}$ and the disorder strength~$D$.

\begin{figure}[h]
\begin{center}
\includegraphics[width=1\linewidth]{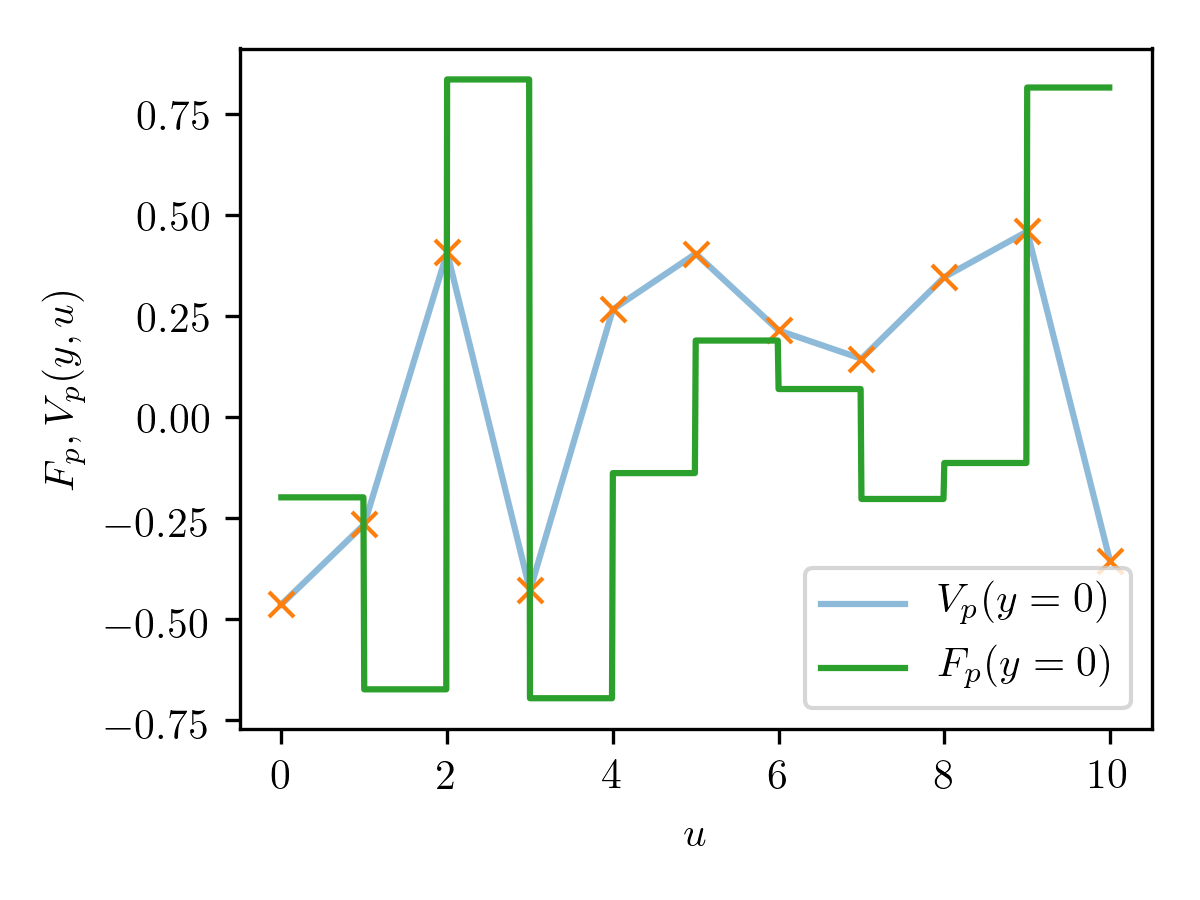}
\caption{
For a fixed coordinate $y$, the random potential ${V_p(y,u)}$ is generated by a linear interpolation between random numbers taken from a uniform distribution with zero mean (orange crosses).
The associated pinning force $F_p$ is obtained from the associated piecewise derivative ${F_p(y,u)=-\partial_u V_p(y,u)}$.
}
\label{fig:UF_correlateddisorder}
\end{center}
\end{figure}

\subsection{Generating a spatially-correlated disorder}
\label{appendix-pinning-force-correlator-disorder-strength-generating}

The pinning potential ${V_p(y,u)}$ is defined independently for each discrete value of the coordinate $y$, so that in what follows, one fixes $y$ and considers only the coordinate $u$.
The procedure to generate the corresponding quenched `landscape' ${U(u)\equiv V_p(y,u)}$ at fixed~$y$, spatially-correlated with a finite correlation length~$\xi$,
is the following (see Fig.~\ref{fig:UF_correlateddisorder}). We first discretize the direction $u$ with fixed steps ${\Delta u=\xi}$.
A random number $U_k$ is generated independently at each site ${u = k \Delta u}$ (with $k$ an integer) from a probability distribution function ${\mathcal{P}(U_k)}$ of zero mean.

Then, on every interval
${u \in [ k \Delta u , (k+1)\Delta u]}$ (\textit{i.e.}~one has ${k = \lfloor u/\Delta u\rfloor}$) the random pinning potential is defined as a linear interpolation between $U_k$ and $U_{k+1}$:
\begin{equation}
\label{eq:defUx}
   U(u) = U_k +  \frac{U_{k+1} - U_k} {\Delta u} ( {u - k \Delta u})
   \, .
\end{equation}
One can easily check that, as required, this definition satisfies
${U(k \Delta u) = U_k}$
and
${U((k+1) \Delta u) = U_{k+1}}$.
The associated pinning force ${F_p(u) = - \partial_u U(u)}$
is finally given by the discrete derivative
\begin{equation}
     F_p(u) = - \frac{U_{k+1} - U_k}{\Delta u}
\label{eq:defFp}
 \end{equation}
with $ {k = \lfloor u/\Delta u\rfloor}$ as above.

For our simulations we sort each reference value $U_k$ from a \emph{uniform} distribution on 
the interval ${[-\frac{\epsilon}{2},\frac{\epsilon}{2}]}$, with ${\epsilon>0}$.
It has consequently a zero mean and a variance
\begin{equation}
	D_0 \equiv \overline{(U_k)^2} = \frac{\epsilon^2}{12}
\, .
\label{eq:D0}
\end{equation}
To keep the discussion general, thereafter we generically denote the variance of $U_k$ as a control parameter $D_0$.
In addition,
we discuss in Appendix~\ref{appendix-disorder-Gaussian-distribution} how choosing an alternative distribution ${\mathcal{P}(U_k)}$ (non-uniform but with the same variance) leads to physically consistent results.

\subsection{Piecewise linear force correlator}
\label{appendix-pinning-force-correlator-disorder-strength-correlator}

We start by defining the intermediate two-point correlator of the force as
\begin{equation}
	\Delta^{(2)}_\xi (u_1,u_2) = \overline{ F_p(u_1) F_p(u_2)}
	\, .
\label{eq:defDeltaxi}
\end{equation}
Because it is associated to the specific set of intervals
${u \in [ k \Delta u , (k+1)\Delta u]}$
with ${k = \lfloor u/\Delta u\rfloor}$,
it is important to notice that it is \emph{not} invariant by translation along the $u$ direction.
Indeed, pairs of points ${(u_1,u_2)}$ separated by a same distance ${u=u_2 - u_1}$ can either lie in the same interval $[k \Delta u,(k+1) \Delta u]$ or not.

One has in fact three possibilities: if ${(u_1,u_2)}$ are 
\begin{itemize}
\item 
in the same interval, one has 
\begin{equation}
   \Delta^{(2)}_\xi (u_1,u_2) = 2 \overline{ (U_i)^2} =  \frac{2D_0}{ \Delta u^2}
	\,;
\label{eq:defDeltaxi-same}
\end{equation}
\item
in adjacent intervals 
$[(k-1) \Delta u, k \Delta u]$ and $[k \Delta u,(k+1) \Delta u]$:
\begin{equation}
	\Delta^{(2)}_\xi (u_1,u_2) = - \overline{ (U_i)^2} = - \frac{D_0}{\Delta u^2}
	\, ;
\label{eq:defDeltaxi-adjacent}
\end{equation}
\item
in more distant intervals
$[k \Delta u,(k+1) \Delta u]$ and $[j \Delta u,(j+1) \Delta u]$ 
with ${|k-j| \leq  2}$:
\begin{equation}
     \Delta^{(2)}_\xi (u_1,u_2) = 0
	\, .
\label{eq:defDeltaxi-distant}
\end{equation}

\end{itemize}
And from now on we use that ${\xi=\Delta u}$ to emphasize the explicit dependence on the correlation length $\xi$.

\begin{figure}[h]
\begin{center}
\includegraphics[width=1\linewidth]{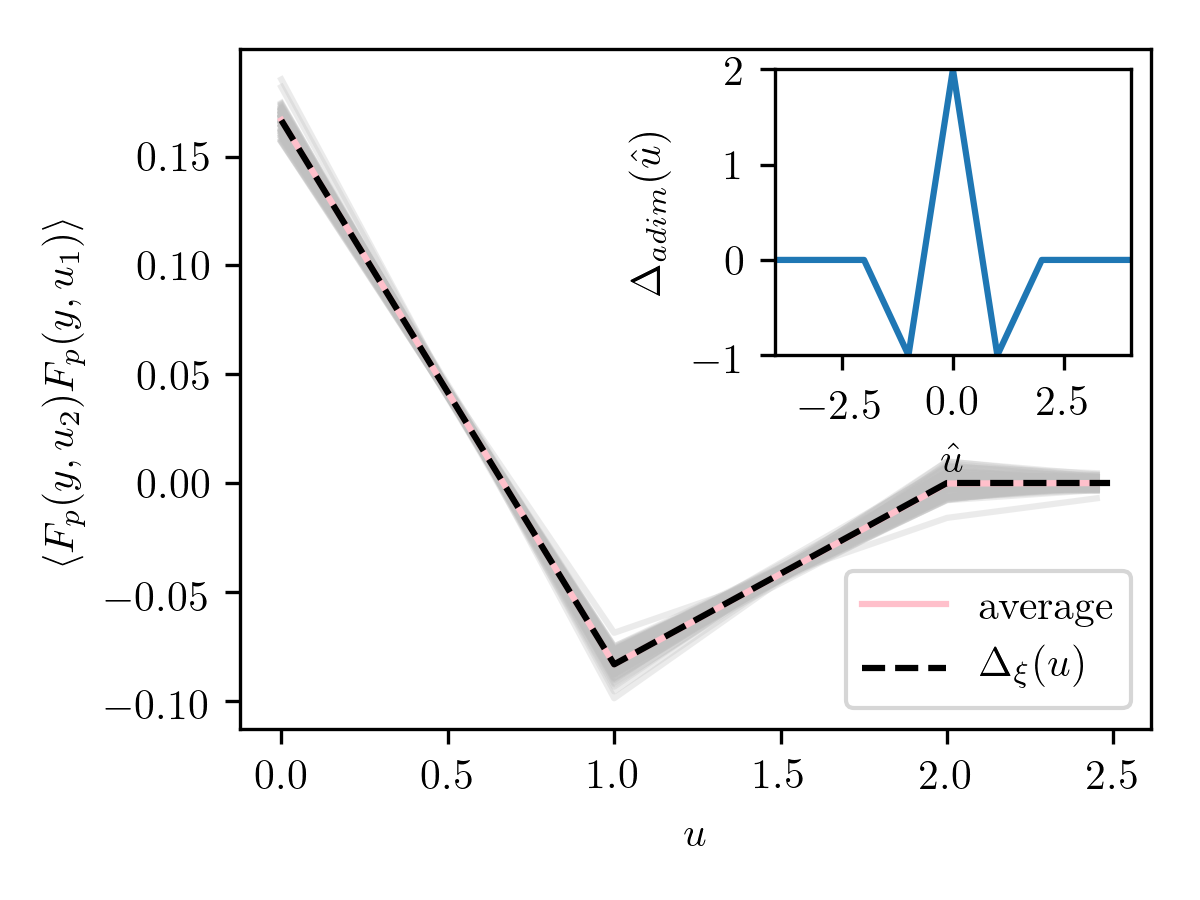}
\caption{
Numerical evaluation of the pinning force correlator for 100 different disorder realizations with $\epsilon=1$ (gray lines) and its average (pink).
In black dashed line we show ${\Delta_\xi(u)}$, the expected correlator given by Eq.~\eqref{eq:Deltaxiu} with $D_0=1/12$. In the inset, we show the adimensionalized force correlator ${\Delta_{\text{adim}}(\hat u)}$ connecting the values of Eqs.~\eqref{eq:Deltaadim1}-\eqref{eq:Deltaadim3}.
}
\label{fig:correlator}
\end{center}
\end{figure}

To recover a translation-invariant correlator, as required in the definitions~\eqref{eq:pinningcorrelations-bis}, one must average the intermediate correlator ${\Delta_\xi^{(2)} (u_1,u_2)}$
over all pairs of points ${(u_1, u_2)}$ separated by the same distance $u$: 
\begin{equation}
	\Delta_\xi (u)
	= \int_{\mathbb{R}^2} \de u_1 \de u_2 \: \delta(u_2-u_1-u) \,\Delta^{(2)}_\xi (u_1,u_2)
\label{eq:force-force-correlator-TI}
\, .
\end{equation}
One finds by an explicit computation
\begin{equation}
	\Delta_\xi(u)
	= \frac{D_0}{\xi^2} \, \Delta_{\text{adim}} \big(u/\xi \big)
\label{eq:Deltaxiu}
\end{equation}
with ${\Delta_{\text{adim}}(\hat u)}$ the piecewise linear continuous function that connects the values:
\begin{align}
\Delta_{\text{adim}}(\hat u)= 2\quad\text{for}\quad\hat u=0 \, ,
\label{eq:Deltaadim1}
\\
\Delta_{\text{adim}}(\hat u) = -1\quad\text{for}\quad|\hat u|=1 \, ,
\\
\Delta_{\text{adim}}(\hat u) = 0 \quad\text{for}\quad|\hat u| \geq 2 \, .
\label{eq:Deltaadim3}
\end{align}
The complete function is plotted in the inset of Fig.~\ref{fig:correlator}.

As a self-consistent validation of our procedure, we evaluated numerically the correlator ${\Delta_\xi(u)}$, and as shown in Fig.~\ref{fig:correlator} we find an excellent agreement with the expression~\eqref{eq:Deltaxiu}.

%

\subsection{Correlators in Fourier space and disorder strength}
\label{appendix-pinning-force-correlator-disorder-strength-Fourier}

One can check that ${\int_{\mathbb{R}} \de \hat{u} \, \Delta_{\text{adim}}(\hat u) = 0}$, as expected for the `random-bond' disorder we consider.
To access the disorder strength $D$, we switch to Fourier space where we can more easily exploit the relation between the correlators ${\Delta_\xi(u)=-D R_{\xi}''(u)}$ from Eq.~\eqref{eq:pinningcorrelations-bis}.

We first rewrite, similarly to Eq.~\eqref{eq:Deltaxiu}, the random potential correlator in terms of its adimensionalized version, starting from its very definition
\begin{equation}
	\overline{U(u) U(0)}
	\equiv D R_\xi(u)
	\equiv \frac {D}{\xi} R_{\text{adim}}(u/\xi)
	\, .
\end{equation}
First, by direct comparison with the definition in Eq.~\eqref{eq:D0} we can establish its relation to the variance $D_0$:
\begin{equation}
	D_0
	\equiv \overline{U(0)^2}
	=D R_\xi(u=0)
	=\frac{D}{\xi} R_{\text{adim}}(\hat{u}=0)
	\, ,
\label{eq:def-D0-D-Rhat}
\end{equation}
and secondly the relation between the correlators ${ \Delta_\xi(u)=-D R_{\xi}''(u)}$ becomes
\begin{equation}
	\frac{D_0}{\xi^2} \Delta_{\text{adim}}(\hat{u}) = - \frac{D}{\xi^3}  R_{\text{adim}}''(\hat{u})
	\, .
\label{eq:relation-Delta-R-adim}
\end{equation}

Defining the Fourier transform along the transverse direction as
${\hat \Delta_{\text{adim}}(\hat q) = \int_{\mathbb{R}} \de \hat u\, e^{i \hat q \hat u} \Delta_{\text{adim}}(\hat u)}$,
one finds by direct computation that it takes the simple form:
\begin{equation}
  \hat \Delta_{\text{adim}}( \hat q)
  	= 16 \frac{\big[ \sin\frac{\hat q}{2}\big]^4}{\hat q^2}
  	\, .
\label{eq:Deltaadim}
\end{equation}
Defining similarly the Fourier transform
${\hat R_{\text{adim}}(\hat q) = \int_{\mathbb{R}} \de \hat u\, e^{i \hat q \hat u} R_{\text{adim}}(\hat u)}$, Eq.~\eqref{eq:relation-Delta-R-adim} rewrites
\begin{equation}
 \hat\Delta_{\text{adim}}(\hat q)
 	= \frac{D}{D_0 \xi} \hat{q}^2 \hat R_{\text{adim}}(\hat q)
 	\, .
\end{equation}
At this point, we might think that we have some freedom to define ${\hat R_{\text{adim}}}$ up to an arbitrary constant, but we have in fact to enforce the imposed normalization
${\int_{\mathbb{R}}\de u \, R_{\text{adim}}(\hat{u})=1}$
or equivalently
${\hat R_{\text{adim}}( \hat{q}=0)=1}$.
This is achieved by imposing ${\frac{D}{D_0 \xi}=1}$ and thus
\begin{equation}
  \hat R_{\text{adim}}( \hat q)
  	= 16 \frac{\big[ \sin\frac{\hat q}{2}\big]^4}{\hat q^4}
  	\quad \Rightarrow \quad
  \lim_{\hat{q} \to 0} \hat R_{\text{adim}}( \hat q) = 1
  \, .
\label{eq:Deltaadim}
\end{equation}
Furthermore, we have by inverse Fourier transform:
\begin{equation}
 R_{\text{adim}}(\hat{u}=0)
 = \int_{\mathbb{R}} \frac{\de \hat{q}}{2 \pi} \, \hat{R}_{\text{adim}} (\hat{q})
 = \frac23
 \, .
\label{eq-linear-Rhat-peak}
\end{equation}
We have, at last, directly access to the disorder strength:
\begin{equation}
 	D
 	\stackrel{\eqref{eq:def-D0-D-Rhat}}{=}
	 	\frac{D_0 \xi}{R_{\text{adim}}(\hat{u}=0)}
	 \stackrel{\eqref{eq-linear-Rhat-peak}}{=}
	 	\frac32 D_0 \xi
	\, .
\end{equation}

The last expression is valid for any random potential generated by a linear interpolation between uncorrelated random points, drawn from an arbitrary  distribution ${\mathcal{P}(U_k)}$ with zero mean and variance $D_0$.
For the uniform distribution we consider, Eq.~\eqref{eq:D0} implies
\begin{equation}
	D^{\text{uniform}} = \frac32 D_0 \xi = \frac18 \epsilon^2
	\, .
\end{equation}
This is the expression that we used at the end of Sec.~\ref{sec-model} in the main text in order to fix the disorder strength.

\section{Parameters of numerical simulations}
\label{appendix-parameters-numerical-simulations}

To solve the quenched Edwards-Wilkinson equation (Eq.~\eqref{eq:qEW} of the main text) and compute the interface roughness, we take advantage of massively parallel accelerated computing with a CUDA C++ code running in NVIDIA GPUs with a Volta architecture, in double precision.
At each simulation step we approximate the second derivative of $u$ along the $y$-direction by a two-point central finite difference scheme and integrate in time with a first-order Euler step.
Pseudo-random numbers are generated with a counter-based RNG (Philox, allowing $2^{64}$ parallel and distinct streams with a period of $2^{128}$~\cite{salmon2011parallel}).
For the thermal noise, we use Gaussian distributed numbers while for the quenched disorder, we use the method described in Appendix~\ref{appendix-pinning-force-correlator-disorder-strength} (and in Appendix~\ref{appendix-disorder-Gaussian-distribution} in the consistency check described in the same section):
the method consists in a linear interpolation of uniformly distributed random numbers with the implementation proposed in~\cite{Ferrero2013nonsteady}. These random numbers, uncorrelated from site to site, are dynamically generated along the evolution of the interface, \textit{i.e.}~we build the disorder at larger $u$ only if the interface has locally wandered further away.

Space discretization along the $y$-direction is set to $1$ and time discretization to $10^{-2}$. We can simulate 4 realizations of systems of 512 sites for $10^8$ steps in approximately 8 hours. 
These scheme and parameters give roughness functions in the clean case which are in very good agreement with the theoretical prediction (Eq.~\eqref{eq:B(r,t)fromFlat} in the main text).
The corresponding roughness of the simulated systems differs from the theoretically predicted values in less than $10^{-4}$ for $r\leq 50$ and less than $10^{-3}$ for larger values of $r$ (both, simulated and predicted roughness functions are shown in Fig.~1 in the main text).
This excellent agreement is a strong consistency check that provides a good support for the validity of the numerical procedure in the disordered case. 



%

\section{Determination of the best value of $\zeta_\dis$}
\label{appendix-determination-best-value-zetadis}

\begin{figure*}
\begin{center}
\includegraphics[width=1\linewidth]{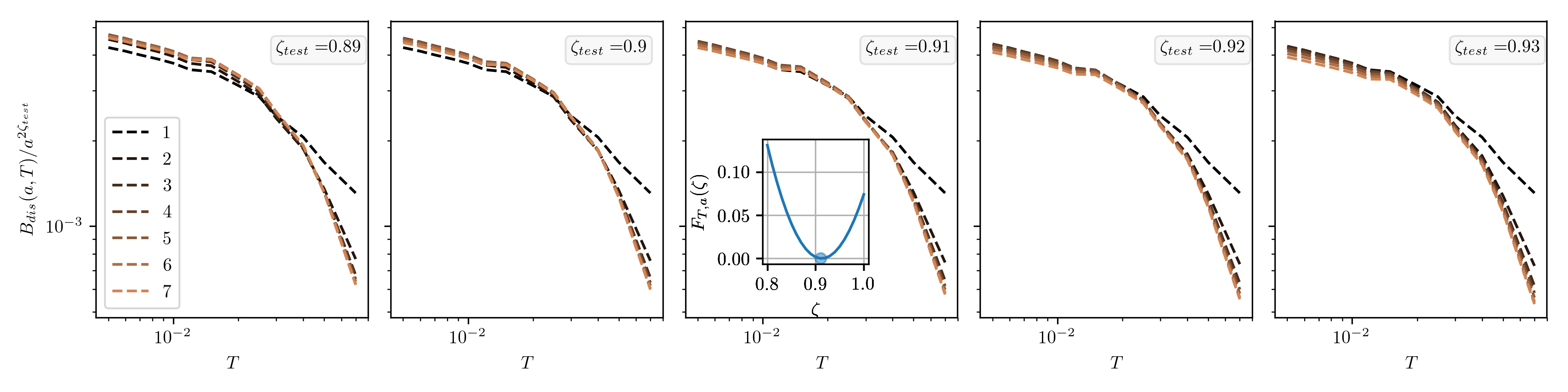}
\caption{Numerical collapse of the excess roughness, described in the text, for
values $\zeta_{\text{test}}$ of $\zeta_\dis$ ranging from  $0.89$ to $0.93$.
The value achieving the best collapse is $\zeta_\dis=0.91$. At 0.91 the function $F_{T,a}(\zeta)$ (see text) has a minimum, as shown in the inset.
The different dashed lines correspond to the value of $a$ indicated in the caption.}
\label{fig:ztest}
\end{center}
\end{figure*}

To obtain $\zeta_\dis$ from the numerical data presented in the main text, we fit $B_\dis(r,T)$ independently for each value of $T$ in the range $r=[1,r_f]$.
For the nine lowest temperatures we have studied, we find values of~$\zeta_\dis$ between 0.9 and 0.92. 
To determine the best value of this exponent that is compatible with every temperature we considered, we rely on the following scaling argument. We predict that $B_\dis(r_0,T)/r_0^{2\zeta_\dis}$ should be independent of $r_0$ for all temperatures $T$.
In Fig.~\ref{fig:ztest}, we illustrate that the best choice of $\zeta_\dis$ that ensures this collapse is $\zeta_\dis=0.91\pm0.01$.
We quantify the spread of the functions $B_\dis(r,T)/r^{2\zeta_{\text{test}}}$ around their mean value for different values of $\zeta_{\text{test}}$ by computing the function $F_{T,r}(\zeta)=\Sigma_{T,r}\Big(\frac{B_\dis(r,T)/r^{2\zeta}-\Sigma_{r_i} B_\dis(r_i,T)/r_i^{2\zeta}}{\Sigma_{r_i} B_\dis(r_i,T)/r_i^{2\zeta}}\Big)^2$ with $r_i=1,\dots,5$, for the 5 lowest studied temperatures. The resulting function has a minimum in $\zeta=0.91$, as shown in the inset of Fig.~\ref{fig:ztest}. 

\begin{figure}
\begin{center}
\includegraphics[width=1\linewidth]{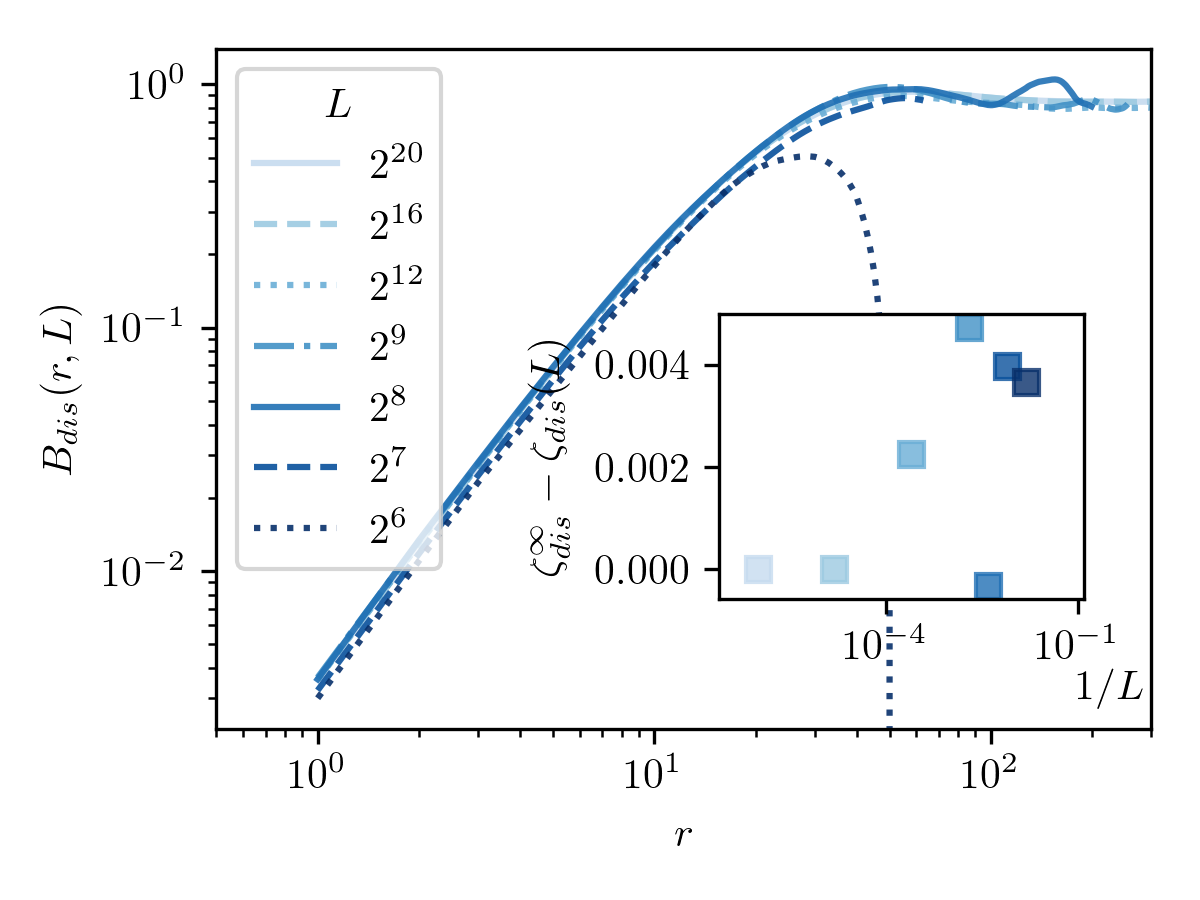}
\caption{Roughness excess obtained for interfaces that evolved for $10^5$ steps under the same conditions at $T=0.01$ for different system sizes $L$ averaged over 50 realizations. In the inset we show the difference between the exponents obtained from fitting the curves with a power-law $\sim\zeta_\dis$ in the range $[1,10]$. $\zeta_\dis^{\infty}$ corresponds to the exponent obtained for the largest system $L=2^{20}$.}
\label{fig:Br_L}
\end{center}
\end{figure}

We note that the regime where we observe the power-law behavior characterized by $\zeta_\dis$ emerges at relatively short scales.  We obtained the roughness of interfaces that evolved from a flat initial condition according to Eq.~\ref{eq:qEW} under the same parameters for different system sizes. As shown in Fig.~\ref{fig:Br_L} at the scales we are interested in the difference in the obtained exponents characterizing the excess power-law regime is of the order $10^{-3}$. For our study we then fix $L=2^9=512$ in this work.

\begin{figure}
\begin{center}
\includegraphics[width=1\linewidth]{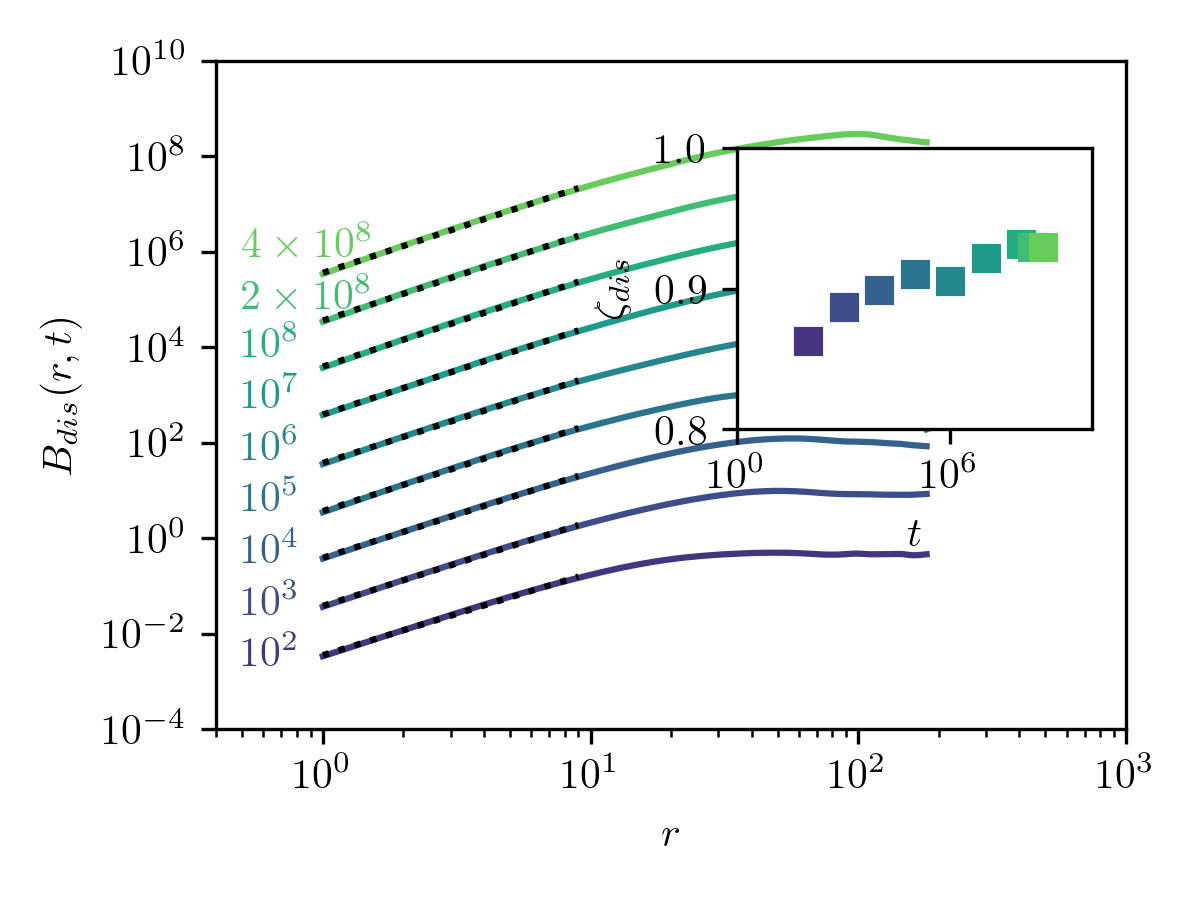}
\caption{Roughness excess obtained for interfaces that evolved for different times (as indicated by the color scale) at $T=0.01$ for a fixed system size $L=512$ averaged over 50 realizations in all cases, except for $t=2\times10^8$, where the average was taken over 10 realizations and $t=4\times10^8$, where the average was taken over 5 realizations. The curves were shifted for clarity. See Fig.~\ref{fig:Bdis_xi2} for the same analysis in the case $\xi=2$.
}
\label{fig:Bdis_t}
\end{center}
\end{figure}

As shown in Fig.~\ref{fig:u_evolution_Br}, the roughness of an initially flat interface will converge to a steady state at sufficiently long times. To test how this influences the regime we are interested in, we follow the roughness of interfaces that evolved for long times at a fixed temperature $T=0.01$. As shown in Fig.~\ref{fig:Bdis_t}, the exponent characterizing the excess regime fluctuates around values which are lower than one.

\begin{figure}
\begin{center}
\includegraphics[width=1\linewidth]{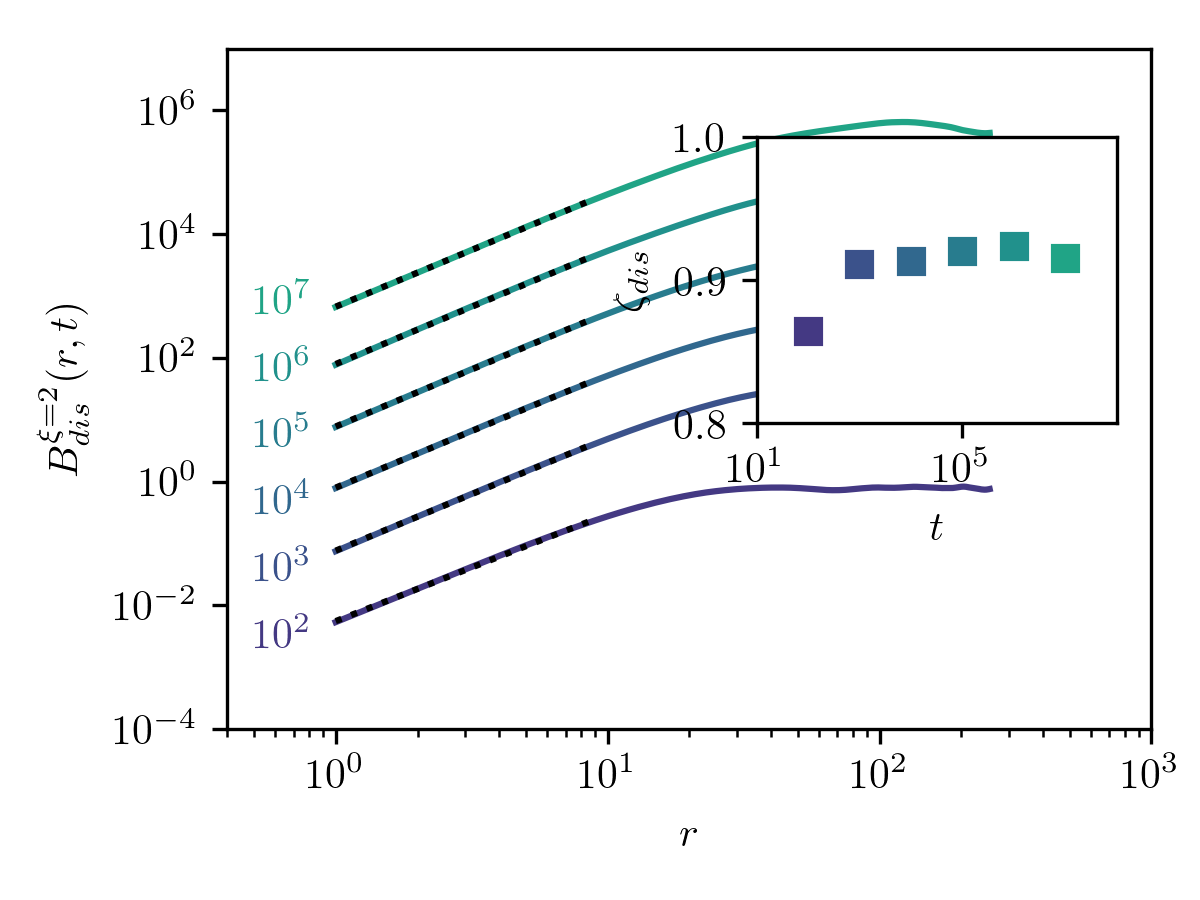}
\caption{Roughness excess obtained for interfaces that evolved for different times (as indicated by the color scale) with a disorder correlation lenght $\xi=2$ at $T=0.01$ for a fixed system size $L=512$ averaged over 50 realizations in all cases, except for the largest studied time, where the average was taken over 7 realizations. The curves were shifted for clarity. In the inset we show the exponents obtained from fitting the curves with a power-law $\sim\zeta_\dis$, shown in dotted black lines. 
}
\label{fig:Bdis_xi2}
\end{center}
\end{figure}

The region in which the excess regime emerges depends on the disorder correlation length $\xi$. We can further verify our numerical findings by studying the roughness excess for interfaces evolving under the same conditions as before, but with a larger disorder correlation length $\xi=2$. We observe that also in this case the exponent $\zeta_\dis$ is compatible with the value 0.91, as shown in Fig.~\ref{fig:Bdis_xi2}.  

\section{Disorder generated from a Gaussian distribution}
\label{appendix-disorder-Gaussian-distribution}

\begin{figure}[t]
\begin{center}
\includegraphics[width=1\linewidth]{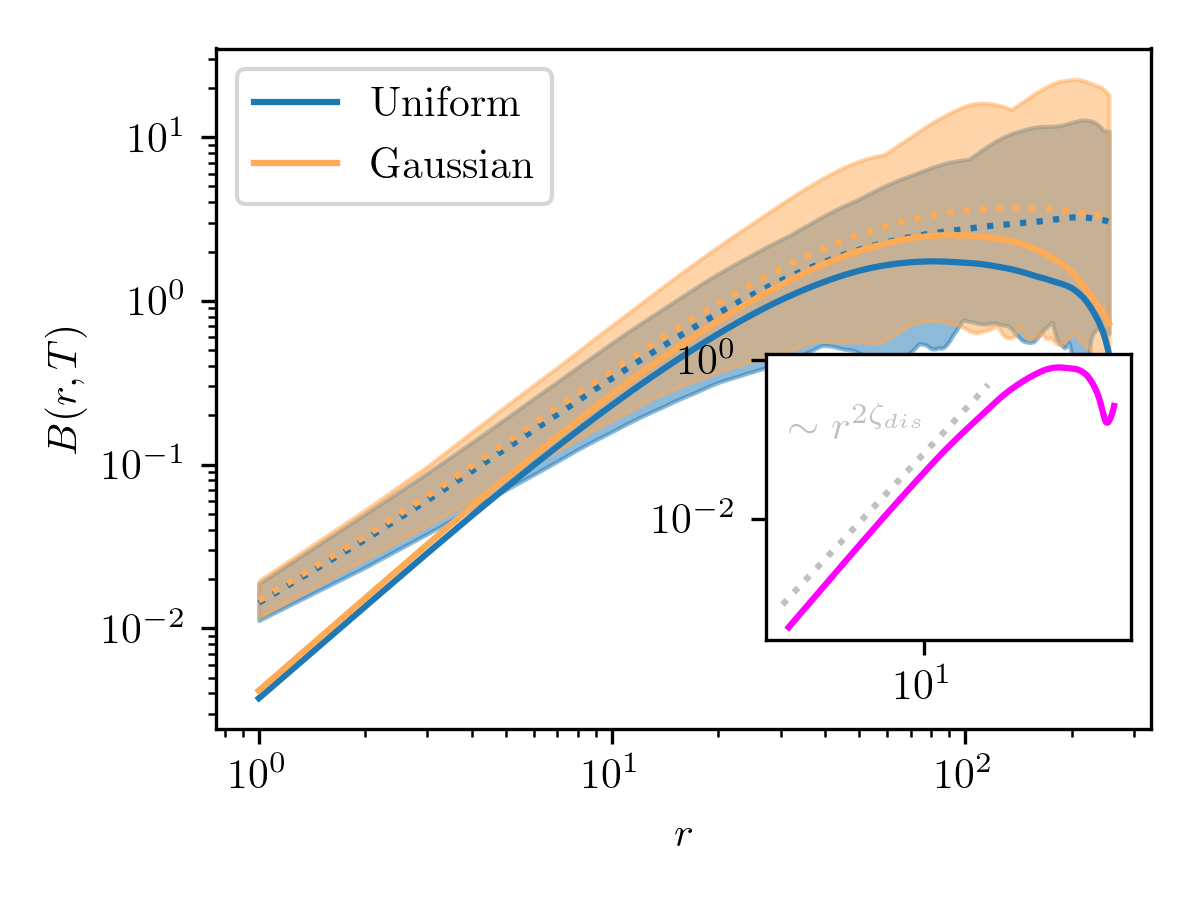}
\caption{Roughness (dotted lines) and excess roughness (continuous lines) at $T=0.01$ averaged over 50 realizations with increased statistics, obtained for disorders generated from a uniform and a Gaussian distribution. In the inset we show the difference between both excess roughness.}
\label{fig:uniform-gaussian}
\end{center}
\end{figure}

For the temperature $T=0.01$, we determined the excess roughness $B_\text{dis}^\text{Gauss}(r)$ for a disorder potential
where the  $U_i$'s are distributed with a Gaussian distribution with the same
variance $\overline{(U_i)^2} = 1/12$ as the uniform one, see Eq.~(\ref{eq:D0}).
On Fig.~\ref{fig:uniform-gaussian}, $B_\text{dis}^\text{Gauss}(r)$ is compared to the excess roughness $B_\dis(r) $ computed
--as in the rest of the paper-- 
for a disordered potential with the $U_i$'s drawn from a uniform distribution, as described in Appendix~\ref{appendix-pinning-force-correlator-disorder-strength}.

The results show that $B_{\text{dis}}^\text{Gauss}(r)$ is very close to $B_\dis(r)$, as mentioned at the end of Sec.~\ref{sec-roughness}.
This provides a strong evidence supporting the following points: in the asymptotic regime $r\to 0$, the power-law behavior $B_\dis(r)\sim r^{2\zeta_\dis} $ is universal, \emph{i.e.~}presents an exponent $\zeta_\dis$ which does not depend on the specific random-potential disorder distribution.
This is an important aspect, as the asymptotic regime $r\to 0$ where $B_\dis(r)$ presents the power-law behavior $\sim r^{2\zeta_\dis}$ could have been sensitive to the details of the disorder correlator at small scales.
Also, the prefactor $A$ in $B_\dis(r)\sim A\,r^{2\zeta_\dis}$ is mainly governed by the variance $D_0$ of the disorder distribution (which is the same in the uniform and in the Gaussian distribution we have used for $\mathcal P(u)$).
Last, we expect also that this prefactor $A$ depends, through a numerical constant, on the rescaled shape $\Delta_{\text{adim}} (\hat u)$ of the disorder correlator. This is seen in the small difference $|B_\dis^\text{Gauss}(r) - B_\dis(r)|\ll B_\dis(r) $ shown in the inset of Fig.~\ref{fig:uniform-gaussian}. Such difference also scales as $r^{2\zeta_\dis}$ in the asymptotic regime $r\to 0$, indicating that indeed only the prefactor $A$ is affected by the adimensionalized shape of the disorder correlator.

\section{Scaling of the crossover scale~$r_1(T)$}
\label{appendix-scaling-crossover}

\begin{figure}[h]
\begin{center}
\includegraphics[width=1\linewidth]{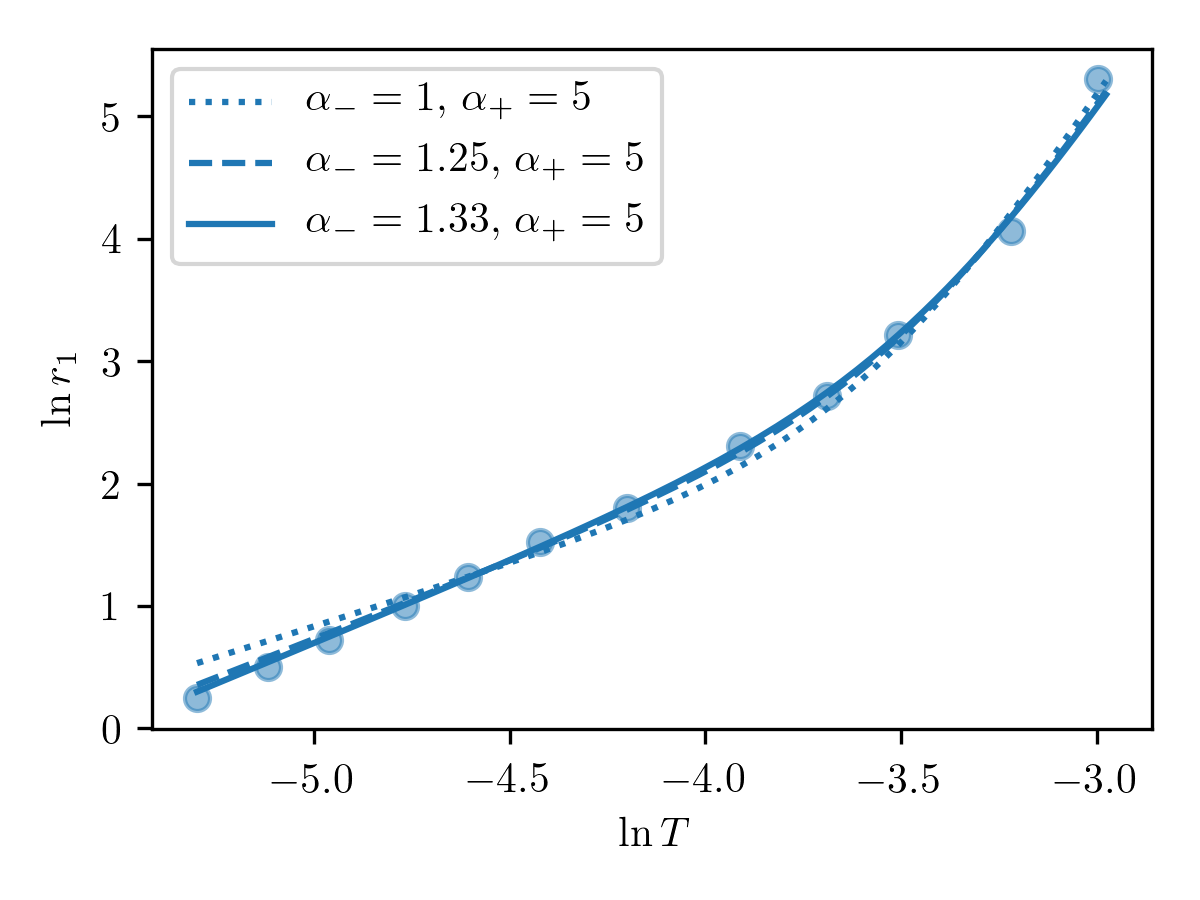}
\caption{Behavior of the crossover scale $r_1(T)$. The points are the result of numerical simulations, while the dashed and full line curves correspond to the two scenarios discussed in the text. The exponent $\alpha_-\approx 1.25 > 1$ is more compatible with the numerical data, and corresponds to $\zeta_\dis\approx 0.9<1$.}
\label{fig:r1crossoverfit}
\end{center}
\end{figure}

At the end of Sec.~\ref{sec-crossovers-scalings}, we mention that the crossover scale $r_1$ is determined numerically by the intersection between the
two power-law scalings ${B_\dis(r) \sim  r^{2\zeta_\dis}}$ and the thermal regime ${T r / c}$.
The scaling arguments presented in the main text indicate that $r_1$ scales with temperature as
\begin{equation}
 r_1(T) \sim \left\lbrace \begin{array}{ll}
 	T^{1/[1-2(1-\zeta_\dis)]}
 	& ~\text{for~} T\ll T_c \, ,
 	\\
     T^5
 	& ~\text{for~} T\gg T_c \, .
 	\end{array} \right.
\label{eq:r1scalingT}
\end{equation}
To determine how our numerical results are compatible with such predictions, we have used a fitting
function 
\begin{equation}
 r_1^\text{fit} (T) = C\, T^{\alpha_-} \big[ 1+ (T/T^\star)^{\alpha_+-\alpha_-} \big]  
\label{eq:r1fit}
\end{equation}
(where $C$ and $T^\star$ are constants and $\alpha_-<\alpha_+$)
that interpolates between the regimes
$r_1^\text{fit}(T)\sim T^{\alpha_-}$ for $T\ll T^\star$ 
and
$r_1^\text{fit}(T)\sim T^{\alpha_+}$ for $T\gg T^\star$.
The prediction of Eq.~\eqref{eq:r1scalingT} corresponds to
\begin{equation}
    \alpha_+=5 \, , \quad \alpha_-=1/[1-2(1-\zeta_\dis)] \, .
\end{equation}
A first scenario where $\zeta_\dis=\zeta_\dis^\text{th}=1$ (according to the perturbative analysis of \cite{korshunov_2013_JETP117_570}, for instance) corresponds to $\alpha_-=1$.
A second scenario where $\zeta_\dis\approx 0.9<1$ corresponds to $\alpha_-\approx 1.25 > 1$.
To distinguish between these two possibilities, we have fitted the values of $r_1(T)$ obtained numerically with the function~\eqref{eq:r1fit}, where we fixed the exponent $\alpha_\pm$ and we left the constants $C$ and $T^\star$ as free parameters.
As shown on~Fig.~\ref{fig:r1crossoverfit}, the data provide a strong evidence supporting the second scenario $\zeta_\dis\approx 0.9<1$.
Also, leaving $\alpha_-$ as a free parameter for the fit, one finds $\alpha_-\approx 1.33$ which corresponds to the value $\zeta_\dis\approx 0.88$: it is compatible with the value $\zeta_\dis\approx 0.91$ obtained in the main text with a completely different method.

\begin{thebibliography}{80}%
\makeatletter
\providecommand \@ifxundefined [1]{%
 \@ifx{#1\undefined}
}%
\providecommand \@ifnum [1]{%
 \ifnum #1\expandafter \@firstoftwo
 \else \expandafter \@secondoftwo
 \fi
}%
\providecommand \@ifx [1]{%
 \ifx #1\expandafter \@firstoftwo
 \else \expandafter \@secondoftwo
 \fi
}%
\providecommand \natexlab [1]{#1}%
\providecommand \enquote  [1]{``#1''}%
\providecommand \bibnamefont  [1]{#1}%
\providecommand \bibfnamefont [1]{#1}%
\providecommand \citenamefont [1]{#1}%
\providecommand \href@noop [0]{\@secondoftwo}%
\providecommand \href [0]{\begingroup \@sanitize@url \@href}%
\providecommand \@href[1]{\@@startlink{#1}\@@href}%
\providecommand \@@href[1]{\endgroup#1\@@endlink}%
\providecommand \@sanitize@url [0]{\catcode `\\12\catcode `\$12\catcode
  `\&12\catcode `\#12\catcode `\^12\catcode `\_12\catcode `\%12\relax}%
\providecommand \@@startlink[1]{}%
\providecommand \@@endlink[0]{}%
\providecommand \url  [0]{\begingroup\@sanitize@url \@url }%
\providecommand \@url [1]{\endgroup\@href {#1}{\urlprefix }}%
\providecommand \urlprefix  [0]{URL }%
\providecommand \Eprint [0]{\href }%
\providecommand \doibase [0]{https://doi.org/}%
\providecommand \selectlanguage [0]{\@gobble}%
\providecommand \bibinfo  [0]{\@secondoftwo}%
\providecommand \bibfield  [0]{\@secondoftwo}%
\providecommand \translation [1]{[#1]}%
\providecommand \BibitemOpen [0]{}%
\providecommand \bibitemStop [0]{}%
\providecommand \bibitemNoStop [0]{.\EOS\space}%
\providecommand \EOS [0]{\spacefactor3000\relax}%
\providecommand \BibitemShut  [1]{\csname bibitem#1\endcsname}%
\let\auto@bib@innerbib\@empty
\bibitem [{\citenamefont {Fisher}(1998)}]{Fisher_review_collective_transport}%
  \BibitemOpen
  \bibfield  {author} {\bibinfo {author} {\bibfnamefont {D.~S.}\ \bibnamefont
  {Fisher}},\ }\href {https://doi.org/DOI: 10.1016/S0370-1573(98)00008-8}
  {\bibfield  {journal} {\bibinfo  {journal} {Phys. Rep.}\ }\textbf {\bibinfo
  {volume} {301}},\ \bibinfo {pages} {113} (\bibinfo {year}
  {1998})}\BibitemShut {NoStop}%
\bibitem [{\citenamefont {Lemerle}\ \emph {et~al.}(1998)\citenamefont
  {Lemerle}, \citenamefont {Ferr{\'e}}, \citenamefont {Chappert}, \citenamefont
  {Mathet}, \citenamefont {Giamarchi},\ and\ \citenamefont {{Le
  Doussal}}}]{lemerle_domainwall_creep}%
  \BibitemOpen
  \bibfield  {author} {\bibinfo {author} {\bibfnamefont {S.}~\bibnamefont
  {Lemerle}}, \bibinfo {author} {\bibfnamefont {J.}~\bibnamefont {Ferr{\'e}}},
  \bibinfo {author} {\bibfnamefont {C.}~\bibnamefont {Chappert}}, \bibinfo
  {author} {\bibfnamefont {V.}~\bibnamefont {Mathet}}, \bibinfo {author}
  {\bibfnamefont {T.}~\bibnamefont {Giamarchi}},\ and\ \bibinfo {author}
  {\bibfnamefont {P.}~\bibnamefont {{Le Doussal}}},\ }\href
  {https://doi.org/10.1103/PhysRevLett.80.849} {\bibfield  {journal} {\bibinfo
  {journal} {Phys. Rev. Lett.}\ }\textbf {\bibinfo {volume} {80}},\ \bibinfo
  {pages} {849} (\bibinfo {year} {1998})}\BibitemShut {NoStop}%
\bibitem [{\citenamefont {Ferr{\'e}}\ \emph {et~al.}(2013)\citenamefont
  {Ferr{\'e}}, \citenamefont {Metaxas}, \citenamefont {Mougin}, \citenamefont
  {Jamet}, \citenamefont {Gorchon},\ and\ \citenamefont
  {Jeudy}}]{ferre_2013_ComptesRendusPhys14_651}%
  \BibitemOpen
  \bibfield  {author} {\bibinfo {author} {\bibfnamefont {J.}~\bibnamefont
  {Ferr{\'e}}}, \bibinfo {author} {\bibfnamefont {P.~J.}\ \bibnamefont
  {Metaxas}}, \bibinfo {author} {\bibfnamefont {A.}~\bibnamefont {Mougin}},
  \bibinfo {author} {\bibfnamefont {J.-P.}\ \bibnamefont {Jamet}}, \bibinfo
  {author} {\bibfnamefont {J.}~\bibnamefont {Gorchon}},\ and\ \bibinfo {author}
  {\bibfnamefont {V.}~\bibnamefont {Jeudy}},\ }\href
  {https://doi.org/10.1016/j.crhy.2013.08.001} {\bibfield  {journal} {\bibinfo
  {journal} {Comptes Rendus Physique}\ }\textbf {\bibinfo {volume} {14}},\
  \bibinfo {pages} {651} (\bibinfo {year} {2013})}\BibitemShut {NoStop}%
\bibitem [{\citenamefont {Paruch}\ and\ \citenamefont
  {Guyonnet}(2013)}]{paruch_2013_ComptesRendusPhys14_667}%
  \BibitemOpen
  \bibfield  {author} {\bibinfo {author} {\bibfnamefont {P.}~\bibnamefont
  {Paruch}}\ and\ \bibinfo {author} {\bibfnamefont {J.}~\bibnamefont
  {Guyonnet}},\ }\href
  {https://doi.org/http://dx.doi.org/10.1016/j.crhy.2013.08.004} {\bibfield
  {journal} {\bibinfo  {journal} {Comptes Rendus Physique}\ }\textbf {\bibinfo
  {volume} {14}},\ \bibinfo {pages} {667 } (\bibinfo {year}
  {2013})}\BibitemShut {NoStop}%
\bibitem [{\citenamefont {Durin}\ \emph {et~al.}(2016)\citenamefont {Durin},
  \citenamefont {Bohn}, \citenamefont {Corr\^ea}, \citenamefont {Sommer},
  \citenamefont {Le~Doussal},\ and\ \citenamefont
  {Wiese}}]{durin_2016_PRL_magneticavalanches}%
  \BibitemOpen
  \bibfield  {author} {\bibinfo {author} {\bibfnamefont {G.}~\bibnamefont
  {Durin}}, \bibinfo {author} {\bibfnamefont {F.}~\bibnamefont {Bohn}},
  \bibinfo {author} {\bibfnamefont {M.~A.}\ \bibnamefont {Corr\^ea}}, \bibinfo
  {author} {\bibfnamefont {R.~L.}\ \bibnamefont {Sommer}}, \bibinfo {author}
  {\bibfnamefont {P.}~\bibnamefont {Le~Doussal}},\ and\ \bibinfo {author}
  {\bibfnamefont {K.~J.}\ \bibnamefont {Wiese}},\ }\href
  {https://doi.org/10.1103/PhysRevLett.117.087201} {\bibfield  {journal}
  {\bibinfo  {journal} {Phys. Rev. Lett.}\ }\textbf {\bibinfo {volume} {117}},\
  \bibinfo {pages} {087201} (\bibinfo {year} {2016})}\BibitemShut {NoStop}%
\bibitem [{\citenamefont {Caballero}\ \emph {et~al.}(2017)\citenamefont
  {Caballero}, \citenamefont {FernandezAguirre}, \citenamefont {Albornoz}, \citenamefont
  {Kolton}, \citenamefont {Rojas-S{\'a}nchez}, \citenamefont {Collin},
  \citenamefont {George}, \citenamefont {Pardo}, \citenamefont {Jeudy},
  \citenamefont {Bustingorry},\ and\ \citenamefont
  {Curiale}}]{caballero2017excess}%
  \BibitemOpen
  \bibfield  {author} {\bibinfo {author} {\bibfnamefont {N.~B.}\ \bibnamefont
  {Caballero}}, \bibinfo {author} {\bibfnamefont {I.}\ \bibnamefont
  {Aguirre}}, \bibinfo {author} {\bibfnamefont {L.~J.}\ \bibnamefont
  {Albornoz}}, \bibinfo {author} {\bibfnamefont {A.~B.}\ \bibnamefont
  {Kolton}}, \bibinfo {author} {\bibfnamefont {J.~C.}\ \bibnamefont
  {Rojas-S{\'a}nchez}}, \bibinfo {author} {\bibfnamefont {S.}~\bibnamefont
  {Collin}}, \bibinfo {author} {\bibfnamefont {J.~M.}\ \bibnamefont {George}},
  \bibinfo {author} {\bibfnamefont {R.~D.}\ \bibnamefont {Pardo}}, \bibinfo
  {author} {\bibfnamefont {V.}~\bibnamefont {Jeudy}}, \bibinfo {author}
  {\bibfnamefont {S.}~\bibnamefont {Bustingorry}},\ and\ \bibinfo {author}
  {\bibfnamefont {J.}~\bibnamefont {Curiale}},\ }\href
  {https://doi.org/10.1103/PhysRevB.96.224422} {\bibfield  {journal} {\bibinfo
  {journal} {Phys. Rev. B}\ }\textbf {\bibinfo {volume} {96}},\ \bibinfo
  {pages} {224422} (\bibinfo {year} {2017})}\BibitemShut {NoStop}%
\bibitem [{\citenamefont {Diaz~Pardo}\ \emph {et~al.}(2017)\citenamefont
  {Diaz~Pardo}, \citenamefont {Savero~Torres}, \citenamefont {Kolton},
  \citenamefont {Bustingorry},\ and\ \citenamefont {Jeudy}}]{pardo2017}%
  \BibitemOpen
  \bibfield  {author} {\bibinfo {author} {\bibfnamefont {R.}~\bibnamefont
  {Diaz~Pardo}}, \bibinfo {author} {\bibfnamefont {W.}~\bibnamefont
  {Savero~Torres}}, \bibinfo {author} {\bibfnamefont {A.~B.}\ \bibnamefont
  {Kolton}}, \bibinfo {author} {\bibfnamefont {S.}~\bibnamefont
  {Bustingorry}},\ and\ \bibinfo {author} {\bibfnamefont {V.}~\bibnamefont
  {Jeudy}},\ }\href
  {https://journals.aps.org/prb/abstract/10.1103/PhysRevB.95.184434} {\bibfield
   {journal} {\bibinfo  {journal} {Phys. Rev. B}\ }\textbf {\bibinfo {volume}
  {95}},\ \bibinfo {pages} {184434} (\bibinfo {year} {2017})}\BibitemShut
  {NoStop}%
\bibitem [{\citenamefont {Salje}\ \emph {et~al.}(2019)\citenamefont {Salje},
  \citenamefont {Xue}, \citenamefont {Ding}, \citenamefont {Dahmen},\ and\
  \citenamefont {Scott}}]{salje_2019_PRM_ferroelectricavalanches}%
  \BibitemOpen
  \bibfield  {author} {\bibinfo {author} {\bibfnamefont {E.~K.~H.}\
  \bibnamefont {Salje}}, \bibinfo {author} {\bibfnamefont {D.}~\bibnamefont
  {Xue}}, \bibinfo {author} {\bibfnamefont {X.}~\bibnamefont {Ding}}, \bibinfo
  {author} {\bibfnamefont {K.~A.}\ \bibnamefont {Dahmen}},\ and\ \bibinfo
  {author} {\bibfnamefont {J.~F.}\ \bibnamefont {Scott}},\ }\href
  {https://doi.org/10.1103/PhysRevMaterials.3.014415} {\bibfield  {journal}
  {\bibinfo  {journal} {Phys. Rev. Materials}\ }\textbf {\bibinfo {volume}
  {3}},\ \bibinfo {pages} {014415} (\bibinfo {year} {2019})}\BibitemShut
  {NoStop}%
\bibitem [{\citenamefont {T{\"u}ckmantel}\ \emph {et~al.}(2021)\citenamefont
  {T{\"u}ckmantel}, \citenamefont {Gaponenko}, \citenamefont {Caballero},
  \citenamefont {Agar}, \citenamefont {Martin}, \citenamefont {Giamarchi},\
  and\ \citenamefont {Paruch}}]{tuckmantel_2021_local}%
  \BibitemOpen
  \bibfield  {author} {\bibinfo {author} {\bibfnamefont {P.}~\bibnamefont
  {T{\"u}ckmantel}}, \bibinfo {author} {\bibfnamefont {I.}~\bibnamefont
  {Gaponenko}}, \bibinfo {author} {\bibfnamefont {N.}~\bibnamefont
  {Caballero}}, \bibinfo {author} {\bibfnamefont {J.~C.}\ \bibnamefont {Agar}},
  \bibinfo {author} {\bibfnamefont {L.~W.}\ \bibnamefont {Martin}}, \bibinfo
  {author} {\bibfnamefont {T.}~\bibnamefont {Giamarchi}},\ and\ \bibinfo
  {author} {\bibfnamefont {P.}~\bibnamefont {Paruch}},\ }\href
  {https://journals.aps.org/prl/abstract/10.1103/PhysRevLett.126.117601}
  {\bibfield  {journal} {\bibinfo  {journal} {Phys. Rev. Lett.}\ }\textbf
  {\bibinfo {volume} {126}},\ \bibinfo {pages} {117601} (\bibinfo {year}
  {2021})}\BibitemShut {NoStop}%
\bibitem [{\citenamefont {Alava}\ \emph {et~al.}(2004)\citenamefont {Alava},
  \citenamefont {Dube},\ and\ \citenamefont {Rost}}]{alava_2004_AdvPhys53_83}%
  \BibitemOpen
  \bibfield  {author} {\bibinfo {author} {\bibfnamefont {M.}~\bibnamefont
  {Alava}}, \bibinfo {author} {\bibfnamefont {M.}~\bibnamefont {Dube}},\ and\
  \bibinfo {author} {\bibfnamefont {M.}~\bibnamefont {Rost}},\ }\href
  {https://doi.org/10.1080/00018730410001687363} {\bibfield  {journal}
  {\bibinfo  {journal} {Advances in Physics}\ }\textbf {\bibinfo {volume}
  {53}},\ \bibinfo {pages} {83} (\bibinfo {year} {2004})}\BibitemShut {NoStop}%
\bibitem [{\citenamefont {Santucci}\ \emph {et~al.}(2007)\citenamefont
  {Santucci}, \citenamefont {M\aa{}l\o{}y}, \citenamefont {Delaplace},
  \citenamefont {Mathiesen}, \citenamefont {Hansen}, \citenamefont
  {Haavig~Bakke}, \citenamefont {Schmittbuhl}, \citenamefont {Vanel},\ and\
  \citenamefont {Ray}}]{santucci_2007_PhysRevE75_016104}%
  \BibitemOpen
  \bibfield  {author} {\bibinfo {author} {\bibfnamefont {S.}~\bibnamefont
  {Santucci}}, \bibinfo {author} {\bibfnamefont {K.~J.}\ \bibnamefont
  {M\aa{}l\o{}y}}, \bibinfo {author} {\bibfnamefont {A.}~\bibnamefont
  {Delaplace}}, \bibinfo {author} {\bibfnamefont {J.}~\bibnamefont
  {Mathiesen}}, \bibinfo {author} {\bibfnamefont {A.}~\bibnamefont {Hansen}},
  \bibinfo {author} {\bibfnamefont {J.~{\O}.}\ \bibnamefont {Haavig~Bakke}},
  \bibinfo {author} {\bibfnamefont {J.}~\bibnamefont {Schmittbuhl}}, \bibinfo
  {author} {\bibfnamefont {L.}~\bibnamefont {Vanel}},\ and\ \bibinfo {author}
  {\bibfnamefont {P.}~\bibnamefont {Ray}},\ }\href
  {http://link.aps.org/doi/10.1103/PhysRevE.75.016104} {\bibfield  {journal}
  {\bibinfo  {journal} {Phys. Rev. E}\ }\textbf {\bibinfo {volume} {75}},\
  \bibinfo {pages} {016104} (\bibinfo {year} {2007})}\BibitemShut {NoStop}%
\bibitem [{\citenamefont {Laurson}\ \emph {et~al.}(2013)\citenamefont
  {Laurson}, \citenamefont {Illa}, \citenamefont {Santucci}, \citenamefont
  {Tallakstad}, \citenamefont {M{\aa}l{\o}y},\ and\ \citenamefont
  {Alava}}]{laurson_NatComm__4_2927_2013cracks}%
  \BibitemOpen
  \bibfield  {author} {\bibinfo {author} {\bibfnamefont {L.}~\bibnamefont
  {Laurson}}, \bibinfo {author} {\bibfnamefont {X.}~\bibnamefont {Illa}},
  \bibinfo {author} {\bibfnamefont {S.}~\bibnamefont {Santucci}}, \bibinfo
  {author} {\bibfnamefont {K.~T.}\ \bibnamefont {Tallakstad}}, \bibinfo
  {author} {\bibfnamefont {K.~J.}\ \bibnamefont {M{\aa}l{\o}y}},\ and\ \bibinfo
  {author} {\bibfnamefont {M.~J.}\ \bibnamefont {Alava}},\ }\href
  {https://www.nature.com/articles/ncomms3927} {\bibfield  {journal} {\bibinfo
  {journal} {Nature Communications}\ }\textbf {\bibinfo {volume} {4}},\
  \bibinfo {pages} {1} (\bibinfo {year} {2013})}\BibitemShut {NoStop}%
\bibitem [{\citenamefont {Santucci}\ \emph {et~al.}(2019)\citenamefont
  {Santucci}, \citenamefont {Tallakstad}, \citenamefont {Angheluta},
  \citenamefont {Laurson}, \citenamefont {Toussaint},\ and\ \citenamefont
  {M{\aa}l{\o}y}}]{santucci_2019_PhilTransRoyalSocA_377_2136_avalanchescrckling}%
  \BibitemOpen
  \bibfield  {author} {\bibinfo {author} {\bibfnamefont {S.}~\bibnamefont
  {Santucci}}, \bibinfo {author} {\bibfnamefont {K.~T.}\ \bibnamefont
  {Tallakstad}}, \bibinfo {author} {\bibfnamefont {L.}~\bibnamefont
  {Angheluta}}, \bibinfo {author} {\bibfnamefont {L.}~\bibnamefont {Laurson}},
  \bibinfo {author} {\bibfnamefont {R.}~\bibnamefont {Toussaint}},\ and\
  \bibinfo {author} {\bibfnamefont {K.~J.}\ \bibnamefont {M{\aa}l{\o}y}},\
  }\href {https://royalsocietypublishing.org/doi/10.1098/rsta.2017.0394}
  {\bibfield  {journal} {\bibinfo  {journal} {Philosophical Transactions of the
  Royal Society A}\ }\textbf {\bibinfo {volume} {377}},\ \bibinfo {pages}
  {20170394} (\bibinfo {year} {2019})}\BibitemShut {NoStop}%
\bibitem [{\citenamefont {Chepizhko}\ \emph {et~al.}(2016)\citenamefont
  {Chepizhko}, \citenamefont {Giampietro}, \citenamefont {Mastrapasqua},
  \citenamefont {Nourazar}, \citenamefont {Ascagni}, \citenamefont {Sugni},
  \citenamefont {Fascio}, \citenamefont {Leggio}, \citenamefont {Malinverno},
  \citenamefont {Scita}, \citenamefont {Santucci}, \citenamefont {Alava},
  \citenamefont {Zapperi},\ and\ \citenamefont
  {La~Porta}}]{chepizhko_2016_PNAS113_11408}%
  \BibitemOpen
  \bibfield  {author} {\bibinfo {author} {\bibfnamefont {O.}~\bibnamefont
  {Chepizhko}}, \bibinfo {author} {\bibfnamefont {C.}~\bibnamefont
  {Giampietro}}, \bibinfo {author} {\bibfnamefont {E.}~\bibnamefont
  {Mastrapasqua}}, \bibinfo {author} {\bibfnamefont {M.}~\bibnamefont
  {Nourazar}}, \bibinfo {author} {\bibfnamefont {M.}~\bibnamefont {Ascagni}},
  \bibinfo {author} {\bibfnamefont {M.}~\bibnamefont {Sugni}}, \bibinfo
  {author} {\bibfnamefont {U.}~\bibnamefont {Fascio}}, \bibinfo {author}
  {\bibfnamefont {L.}~\bibnamefont {Leggio}}, \bibinfo {author} {\bibfnamefont
  {C.}~\bibnamefont {Malinverno}}, \bibinfo {author} {\bibfnamefont
  {G.}~\bibnamefont {Scita}}, \bibinfo {author} {\bibfnamefont
  {S.}~\bibnamefont {Santucci}}, \bibinfo {author} {\bibfnamefont {M.~J.}\
  \bibnamefont {Alava}}, \bibinfo {author} {\bibfnamefont {S.}~\bibnamefont
  {Zapperi}},\ and\ \bibinfo {author} {\bibfnamefont {C.~A.~M.}\ \bibnamefont
  {La~Porta}},\ }\href {https://www.pnas.org/content/113/41/11408} {\bibfield
  {journal} {\bibinfo  {journal} {Proceedings of the National Academy of
  Sciences}\ }\textbf {\bibinfo {volume} {113}},\ \bibinfo {pages} {11408}
  (\bibinfo {year} {2016})}\BibitemShut {NoStop}%
\bibitem [{\citenamefont {Alert}\ and\ \citenamefont
  {Trepat}(2020)}]{alert_2020_AnnRev_collectvecellmigration}%
  \BibitemOpen
  \bibfield  {author} {\bibinfo {author} {\bibfnamefont {R.}~\bibnamefont
  {Alert}}\ and\ \bibinfo {author} {\bibfnamefont {X.}~\bibnamefont {Trepat}},\
  }\href {https://doi.org/10.1146/annurev-conmatphys-031218-013516} {\bibfield
  {journal} {\bibinfo  {journal} {Annual Review of Condensed Matter Physics}\
  }\textbf {\bibinfo {volume} {11}},\ \bibinfo {pages} {77} (\bibinfo {year}
  {2020})}\BibitemShut {NoStop}%
\bibitem [{\citenamefont {Rapin*}\ \emph {et~al.}(2021)\citenamefont {Rapin*},
  \citenamefont {Caballero*}, \citenamefont {Gaponenko}, \citenamefont
  {Ziegler}, \citenamefont {Rawleight}, \citenamefont {Moriggi}, \citenamefont
  {Giamarchi}, \citenamefont {Brown},\ and\ \citenamefont
  {Paruch}}]{rapin_2021_roughness}%
  \BibitemOpen
  \bibfield  {author} {\bibinfo {author} {\bibfnamefont {G.}~\bibnamefont
  {Rapin*}}, \bibinfo {author} {\bibfnamefont {N.}~\bibnamefont {Caballero*}},
  \bibinfo {author} {\bibfnamefont {I.}~\bibnamefont {Gaponenko}}, \bibinfo
  {author} {\bibfnamefont {B.}~\bibnamefont {Ziegler}}, \bibinfo {author}
  {\bibfnamefont {A.}~\bibnamefont {Rawleight}}, \bibinfo {author}
  {\bibfnamefont {E.}~\bibnamefont {Moriggi}}, \bibinfo {author} {\bibfnamefont
  {T.}~\bibnamefont {Giamarchi}}, \bibinfo {author} {\bibfnamefont {S.~A.}\
  \bibnamefont {Brown}},\ and\ \bibinfo {author} {\bibfnamefont
  {P.}~\bibnamefont {Paruch}},\ }\href
  {https://doi.org/10.1038/s41598-021-86684-3} {\bibfield  {journal} {\bibinfo
  {journal} {Scientific Reports}\ }\textbf {\bibinfo {volume} {11}},\ \bibinfo
  {pages} {1} (\bibinfo {year} {2021})}\BibitemShut {NoStop}%
\bibitem [{\citenamefont {Berthier}\ and\ \citenamefont
  {Biroli}(2011)}]{berthier_biroli_2011_RevModPhys83_587}%
  \BibitemOpen
  \bibfield  {author} {\bibinfo {author} {\bibfnamefont {L.}~\bibnamefont
  {Berthier}}\ and\ \bibinfo {author} {\bibfnamefont {G.}~\bibnamefont
  {Biroli}},\ }\href {http://link.aps.org/doi/10.1103/RevModPhys.83.587}
  {\bibfield  {journal} {\bibinfo  {journal} {Rev. Mod. Phys.}\ }\textbf
  {\bibinfo {volume} {83}},\ \bibinfo {pages} {587} (\bibinfo {year}
  {2011})}\BibitemShut {NoStop}%
\bibitem [{\citenamefont {Fisher}(1985)}]{fisher_depinning_meanfield}%
  \BibitemOpen
  \bibfield  {author} {\bibinfo {author} {\bibfnamefont {D.~S.}\ \bibnamefont
  {Fisher}},\ }\href {https://doi.org/10.1103/PhysRevB.31.1396} {\bibfield
  {journal} {\bibinfo  {journal} {Phys. Rev. B}\ }\textbf {\bibinfo {volume}
  {31}},\ \bibinfo {pages} {1396} (\bibinfo {year} {1985})}\BibitemShut
  {NoStop}%
\bibitem [{\citenamefont {Blatter}\ \emph {et~al.}(1994)\citenamefont
  {Blatter}, \citenamefont {Feigel'man}, \citenamefont {Geshkenbein},
  \citenamefont {Larkin},\ and\ \citenamefont
  {Vinokur}}]{blatter_vortex_review}%
  \BibitemOpen
  \bibfield  {author} {\bibinfo {author} {\bibfnamefont {G.}~\bibnamefont
  {Blatter}}, \bibinfo {author} {\bibfnamefont {M.~V.}\ \bibnamefont
  {Feigel'man}}, \bibinfo {author} {\bibfnamefont {V.~B.}\ \bibnamefont
  {Geshkenbein}}, \bibinfo {author} {\bibfnamefont {A.~I.}\ \bibnamefont
  {Larkin}},\ and\ \bibinfo {author} {\bibfnamefont {V.~M.}\ \bibnamefont
  {Vinokur}},\ }\href
  {https://journals.aps.org/rmp/abstract/10.1103/RevModPhys.66.1125} {\bibfield
   {journal} {\bibinfo  {journal} {Rev. Mod. Phys.}\ }\textbf {\bibinfo
  {volume} {66}},\ \bibinfo {pages} {1125} (\bibinfo {year}
  {1994})}\BibitemShut {NoStop}%
\bibitem [{\citenamefont {Chauve}\ \emph {et~al.}(1998)\citenamefont {Chauve},
  \citenamefont {Giamarchi},\ and\ \citenamefont {{Le
  Doussal}}}]{chauve_creep_short}%
  \BibitemOpen
  \bibfield  {author} {\bibinfo {author} {\bibfnamefont {P.}~\bibnamefont
  {Chauve}}, \bibinfo {author} {\bibfnamefont {T.}~\bibnamefont {Giamarchi}},\
  and\ \bibinfo {author} {\bibfnamefont {P.}~\bibnamefont {{Le Doussal}}},\
  }\href {https://doi.org/10.1209/epl/i1998-00443-7} {\bibfield  {journal}
  {\bibinfo  {journal} {Europhys. Lett.}\ }\textbf {\bibinfo {volume} {44}},\
  \bibinfo {pages} {110} (\bibinfo {year} {1998})}\BibitemShut {NoStop}%
\bibitem [{\citenamefont {Chauve}\ \emph {et~al.}(2000)\citenamefont {Chauve},
  \citenamefont {Giamarchi},\ and\ \citenamefont {{Le
  Doussal}}}]{chauve_2000_ThesePC_PhysRevB62_6241}%
  \BibitemOpen
  \bibfield  {author} {\bibinfo {author} {\bibfnamefont {P.}~\bibnamefont
  {Chauve}}, \bibinfo {author} {\bibfnamefont {T.}~\bibnamefont {Giamarchi}},\
  and\ \bibinfo {author} {\bibfnamefont {P.}~\bibnamefont {{Le Doussal}}},\
  }\href {https://doi.org/10.1103/PhysRevB.62.6241} {\bibfield  {journal}
  {\bibinfo  {journal} {Phys. Rev. B}\ }\textbf {\bibinfo {volume} {62}},\
  \bibinfo {pages} {6241} (\bibinfo {year} {2000})}\BibitemShut {NoStop}%
\bibitem [{\citenamefont {Agoritsas}\ \emph
  {et~al.}(2012{\natexlab{a}})\citenamefont {Agoritsas}, \citenamefont
  {Lecomte},\ and\ \citenamefont {Giamarchi}}]{agoritsas_2012_ECRYS2011}%
  \BibitemOpen
  \bibfield  {author} {\bibinfo {author} {\bibfnamefont {E.}~\bibnamefont
  {Agoritsas}}, \bibinfo {author} {\bibfnamefont {V.}~\bibnamefont {Lecomte}},\
  and\ \bibinfo {author} {\bibfnamefont {T.}~\bibnamefont {Giamarchi}},\ }\href
  {https://doi.org/10.1016/j.physb.2012.01.017} {\bibfield  {journal} {\bibinfo
   {journal} {Physica B}\ }\textbf {\bibinfo {volume} {407}},\ \bibinfo {pages}
  {1725} (\bibinfo {year} {2012}{\natexlab{a}})}\BibitemShut {NoStop}%
\bibitem [{\citenamefont {Wiese}(2021)}]{wiese_2021_Arxiv-2102.01215}%
  \BibitemOpen
  \bibfield  {author} {\bibinfo {author} {\bibfnamefont {K.~J.}\ \bibnamefont
  {Wiese}},\ }\href {https://arxiv.org/abs/2102.01215} {\bibinfo {title}
  {Theory and experiments for disordered elastic manifolds, depinning,
  avalanches, and sandpiles}},\ \bibinfo {howpublished} {arXiv:2102.01215
  [cond-mat.dis-nn]} (\bibinfo {year} {2021})\BibitemShut {NoStop}%
\bibitem [{\citenamefont {Ferrero}\ \emph {et~al.}(2021)\citenamefont
  {Ferrero}, \citenamefont {Foini}, \citenamefont {Giamarchi}, \citenamefont
  {Kolton},\ and\ \citenamefont
  {Rosso}}]{ferrero_2021_AnnualReviewsCondMattPhys_creep}%
  \BibitemOpen
  \bibfield  {author} {\bibinfo {author} {\bibfnamefont {E.~E.}\ \bibnamefont
  {Ferrero}}, \bibinfo {author} {\bibfnamefont {L.}~\bibnamefont {Foini}},
  \bibinfo {author} {\bibfnamefont {T.}~\bibnamefont {Giamarchi}}, \bibinfo
  {author} {\bibfnamefont {A.~B.}\ \bibnamefont {Kolton}},\ and\ \bibinfo
  {author} {\bibfnamefont {A.}~\bibnamefont {Rosso}},\ }\href
  {https://www.annualreviews.org/doi/full/10.1146/annurev-conmatphys-031119-050725}
  {\bibfield  {journal} {\bibinfo  {journal} {Annual Review of Condensed Matter
  Physics}\ }\textbf {\bibinfo {volume} {12}},\ \bibinfo {pages} {111}
  (\bibinfo {year} {2021})}\BibitemShut {NoStop}%
\bibitem [{\citenamefont {Barab{\'a}si}\ and\ \citenamefont
  {Stanley}(1995)}]{Barabasi-Stanley}%
  \BibitemOpen
  \bibfield  {author} {\bibinfo {author} {\bibfnamefont {A.-L.}\ \bibnamefont
  {Barab{\'a}si}}\ and\ \bibinfo {author} {\bibfnamefont {H.~E.}\ \bibnamefont
  {Stanley}},\ }\href@noop {} {\emph {\bibinfo {title} {Fractal Concepts in
  Surface Growth}}},\ \bibinfo {edition} {{Cambridge University Press}}\ ed.\
  (\bibinfo {address} {Cambridge},\ \bibinfo {year} {1995})\BibitemShut
  {NoStop}%
\bibitem [{\citenamefont {Jord\'an}\ \emph {et~al.}(2020)\citenamefont
  {Jord\'an}, \citenamefont {Albornoz}, \citenamefont {Gorchon}, \citenamefont
  {Lambert}, \citenamefont {Salahuddin}, \citenamefont {Bokor}, \citenamefont
  {Curiale},\ and\ \citenamefont
  {Bustingorry}}]{jordan_2020_PhysRevB101_184431}%
  \BibitemOpen
  \bibfield  {author} {\bibinfo {author} {\bibfnamefont {D.}~\bibnamefont
  {Jord\'an}}, \bibinfo {author} {\bibfnamefont {L.~J.}\ \bibnamefont
  {Albornoz}}, \bibinfo {author} {\bibfnamefont {J.}~\bibnamefont {Gorchon}},
  \bibinfo {author} {\bibfnamefont {C.~H.}\ \bibnamefont {Lambert}}, \bibinfo
  {author} {\bibfnamefont {S.}~\bibnamefont {Salahuddin}}, \bibinfo {author}
  {\bibfnamefont {J.}~\bibnamefont {Bokor}}, \bibinfo {author} {\bibfnamefont
  {J.}~\bibnamefont {Curiale}},\ and\ \bibinfo {author} {\bibfnamefont
  {S.}~\bibnamefont {Bustingorry}},\ }\href
  {https://doi.org/https://doi.org/10.1103/PhysRevB.101.184431} {\bibfield
  {journal} {\bibinfo  {journal} {Phys. Rev. B}\ }\textbf {\bibinfo {volume}
  {101}},\ \bibinfo {pages} {184431} (\bibinfo {year} {2020})}\BibitemShut
  {NoStop}%
\bibitem [{\citenamefont {Cort\'es~Burgos}\ \emph {et~al.}(2021)\citenamefont
  {Cort\'es~Burgos}, \citenamefont {Guruciaga}, \citenamefont {Jord\'an},
  \citenamefont {Quinteros}, \citenamefont {Agoritsas}, \citenamefont
  {Curiale}, \citenamefont {Granada},\ and\ \citenamefont
  {Bustingorry}}]{cortes_2021_PhysRevB104_144202}%
  \BibitemOpen
  \bibfield  {author} {\bibinfo {author} {\bibfnamefont {M.~J.}\ \bibnamefont
  {Cort\'es~Burgos}}, \bibinfo {author} {\bibfnamefont {P.~C.}\ \bibnamefont
  {Guruciaga}}, \bibinfo {author} {\bibfnamefont {D.}~\bibnamefont {Jord\'an}},
  \bibinfo {author} {\bibfnamefont {C.~P.}\ \bibnamefont {Quinteros}}, \bibinfo
  {author} {\bibfnamefont {E.}~\bibnamefont {Agoritsas}}, \bibinfo {author}
  {\bibfnamefont {J.}~\bibnamefont {Curiale}}, \bibinfo {author} {\bibfnamefont
  {M.}~\bibnamefont {Granada}},\ and\ \bibinfo {author} {\bibfnamefont
  {S.}~\bibnamefont {Bustingorry}},\ }\href
  {https://link.aps.org/doi/10.1103/PhysRevB.104.144202} {\bibfield  {journal}
  {\bibinfo  {journal} {Phys. Rev. B}\ }\textbf {\bibinfo {volume} {104}},\
  \bibinfo {pages} {144202} (\bibinfo {year} {2021})}\BibitemShut {NoStop}%
\bibitem [{\citenamefont {Rapin}\ \emph {et~al.}(2021)\citenamefont {Rapin},
  \citenamefont {Ehrensperger}, \citenamefont {Blaser}, \citenamefont
  {Caballero},\ and\ \citenamefont {Paruch}}]{rapin_2021_APL_dynamicresponse}%
  \BibitemOpen
  \bibfield  {author} {\bibinfo {author} {\bibfnamefont {G.}~\bibnamefont
  {Rapin}}, \bibinfo {author} {\bibfnamefont {S.}~\bibnamefont {Ehrensperger}},
  \bibinfo {author} {\bibfnamefont {C.}~\bibnamefont {Blaser}}, \bibinfo
  {author} {\bibfnamefont {N.}~\bibnamefont {Caballero}},\ and\ \bibinfo
  {author} {\bibfnamefont {P.}~\bibnamefont {Paruch}},\ }\href
  {https://doi.org/10.1063/5.0069920} {\bibfield  {journal} {\bibinfo
  {journal} {Applied Physics Letters}\ }\textbf {\bibinfo {volume} {119}},\
  \bibinfo {pages} {242903} (\bibinfo {year} {2021})}\BibitemShut {NoStop}%
\bibitem [{\citenamefont {Agoritsas}\ \emph {et~al.}(2010)\citenamefont
  {Agoritsas}, \citenamefont {Lecomte},\ and\ \citenamefont
  {Giamarchi}}]{agoritsas_2010_PhysRevB_82_184207}%
  \BibitemOpen
  \bibfield  {author} {\bibinfo {author} {\bibfnamefont {E.}~\bibnamefont
  {Agoritsas}}, \bibinfo {author} {\bibfnamefont {V.}~\bibnamefont {Lecomte}},\
  and\ \bibinfo {author} {\bibfnamefont {T.}~\bibnamefont {Giamarchi}},\ }\href
  {https://doi.org/10.1103/PhysRevB.82.184207} {\bibfield  {journal} {\bibinfo
  {journal} {Phys. Rev. B}\ }\textbf {\bibinfo {volume} {82}},\ \bibinfo
  {pages} {184207} (\bibinfo {year} {2010})}\BibitemShut {NoStop}%
\bibitem [{\citenamefont {Agoritsas}\ \emph
  {et~al.}(2013{\natexlab{a}})\citenamefont {Agoritsas}, \citenamefont
  {Lecomte},\ and\ \citenamefont
  {Giamarchi}}]{agoritsas_2012_FHHtri-analytics}%
  \BibitemOpen
  \bibfield  {author} {\bibinfo {author} {\bibfnamefont {E.}~\bibnamefont
  {Agoritsas}}, \bibinfo {author} {\bibfnamefont {V.}~\bibnamefont {Lecomte}},\
  and\ \bibinfo {author} {\bibfnamefont {T.}~\bibnamefont {Giamarchi}},\ }\href
  {https://doi.org/10.1103/PhysRevE.87.042406} {\bibfield  {journal} {\bibinfo
  {journal} {Phys. Rev. E}\ }\textbf {\bibinfo {volume} {87}},\ \bibinfo
  {pages} {042406} (\bibinfo {year} {2013}{\natexlab{a}})}\BibitemShut
  {NoStop}%
\bibitem [{\citenamefont {Agoritsas}\ \emph
  {et~al.}(2013{\natexlab{b}})\citenamefont {Agoritsas}, \citenamefont
  {Lecomte},\ and\ \citenamefont {Giamarchi}}]{agoritsas_2012_FHHtri-numerics}%
  \BibitemOpen
  \bibfield  {author} {\bibinfo {author} {\bibfnamefont {E.}~\bibnamefont
  {Agoritsas}}, \bibinfo {author} {\bibfnamefont {V.}~\bibnamefont {Lecomte}},\
  and\ \bibinfo {author} {\bibfnamefont {T.}~\bibnamefont {Giamarchi}},\ }\href
  {https://doi.org/10.1103/PhysRevE.87.062405} {\bibfield  {journal} {\bibinfo
  {journal} {Phys. Rev. E}\ }\textbf {\bibinfo {volume} {87}},\ \bibinfo
  {pages} {062405} (\bibinfo {year} {2013}{\natexlab{b}})}\BibitemShut
  {NoStop}%
\bibitem [{\citenamefont {Korshunov}\ \emph {et~al.}(2013)\citenamefont
  {Korshunov}, \citenamefont {Geshkenbein},\ and\ \citenamefont
  {Blatter}}]{korshunov_2013_JETP117_570}%
  \BibitemOpen
  \bibfield  {author} {\bibinfo {author} {\bibfnamefont {S.~E.}\ \bibnamefont
  {Korshunov}}, \bibinfo {author} {\bibfnamefont {V.~B.}\ \bibnamefont
  {Geshkenbein}},\ and\ \bibinfo {author} {\bibfnamefont {G.}~\bibnamefont
  {Blatter}},\ }\href {https://doi.org/10.1134/S1063776113110022} {\bibfield
  {journal} {\bibinfo  {journal} {Journal of Experimental and Theoretical
  Physics}\ }\textbf {\bibinfo {volume} {117}},\ \bibinfo {pages} {570}
  (\bibinfo {year} {2013})}\BibitemShut {NoStop}%
\bibitem [{\citenamefont {Kardar}\ \emph {et~al.}(1986)\citenamefont {Kardar},
  \citenamefont {Parisi},\ and\ \citenamefont
  {Zhang}}]{kardar_1986_originalKPZ_PhysRevLett56_889}%
  \BibitemOpen
  \bibfield  {author} {\bibinfo {author} {\bibfnamefont {M.}~\bibnamefont
  {Kardar}}, \bibinfo {author} {\bibfnamefont {G.}~\bibnamefont {Parisi}},\
  and\ \bibinfo {author} {\bibfnamefont {Y.-C.}\ \bibnamefont {Zhang}},\ }\href
  {https://doi.org/10.1103/PhysRevLett.56.889} {\bibfield  {journal} {\bibinfo
  {journal} {Phys. Rev. Lett.}\ }\textbf {\bibinfo {volume} {56}},\ \bibinfo
  {pages} {889} (\bibinfo {year} {1986})}\BibitemShut {NoStop}%
\bibitem [{\citenamefont {Huse}\ \emph {et~al.}(1985)\citenamefont {Huse},
  \citenamefont {Henley},\ and\ \citenamefont
  {Fisher}}]{huse_henley_fisher_1985_PhysRevLett55_2924}%
  \BibitemOpen
  \bibfield  {author} {\bibinfo {author} {\bibfnamefont {D.~A.}\ \bibnamefont
  {Huse}}, \bibinfo {author} {\bibfnamefont {C.~L.}\ \bibnamefont {Henley}},\
  and\ \bibinfo {author} {\bibfnamefont {D.~S.}\ \bibnamefont {Fisher}},\
  }\href {https://doi.org/10.1103/PhysRevLett.55.2924} {\bibfield  {journal}
  {\bibinfo  {journal} {Phys. Rev. Lett.}\ }\textbf {\bibinfo {volume} {55}},\
  \bibinfo {pages} {2924} (\bibinfo {year} {1985})}\BibitemShut {NoStop}%
\bibitem [{\citenamefont {Corwin}(2012)}]{corwin_2011_arXiv:1106.1596}%
  \BibitemOpen
  \bibfield  {author} {\bibinfo {author} {\bibfnamefont {I.}~\bibnamefont
  {Corwin}},\ }\href {https://doi.org/10.1142/S2010326311300014} {\bibfield
  {journal} {\bibinfo  {journal} {Random Matrices: Theory and Applications}\
  }\textbf {\bibinfo {volume} {01}},\ \bibinfo {pages} {1130001} (\bibinfo
  {year} {2012})}\BibitemShut {NoStop}%
\bibitem [{\citenamefont {Calabrese}\ and\ \citenamefont
  {Le~Doussal}(2011)}]{calabrese_exact_2011}%
  \BibitemOpen
  \bibfield  {author} {\bibinfo {author} {\bibfnamefont {P.}~\bibnamefont
  {Calabrese}}\ and\ \bibinfo {author} {\bibfnamefont {P.}~\bibnamefont
  {Le~Doussal}},\ }\href {https://doi.org/10.1103/PhysRevLett.106.250603}
  {\bibfield  {journal} {\bibinfo  {journal} {Phys. Rev. Lett.}\ }\textbf
  {\bibinfo {volume} {106}},\ \bibinfo {pages} {250603} (\bibinfo {year}
  {2011})}\BibitemShut {NoStop}%
\bibitem [{\citenamefont {Doussal}\ and\ \citenamefont
  {Calabrese}(2012)}]{doussal_kpz_2012}%
  \BibitemOpen
  \bibfield  {author} {\bibinfo {author} {\bibfnamefont {P.~L.}\ \bibnamefont
  {Doussal}}\ and\ \bibinfo {author} {\bibfnamefont {P.}~\bibnamefont
  {Calabrese}},\ }\href {https://doi.org/10.1088/1742-5468/2012/06/P06001}
  {\bibfield  {journal} {\bibinfo  {journal} {J. Stat. Mech.}\ }\textbf
  {\bibinfo {volume} {2012}},\ \bibinfo {pages} {P06001} (\bibinfo {year}
  {2012})}\BibitemShut {NoStop}%
\bibitem [{\citenamefont {Halpin-Healy}\ and\ \citenamefont
  {Takeuchi}(2015)}]{halpin-healy_takeuchi_2015_JStatPhys160_794}%
  \BibitemOpen
  \bibfield  {author} {\bibinfo {author} {\bibfnamefont {T.}~\bibnamefont
  {Halpin-Healy}}\ and\ \bibinfo {author} {\bibfnamefont {K.~A.}\ \bibnamefont
  {Takeuchi}},\ }\href {https://doi.org/10.1007/s10955-015-1282-1} {\bibfield
  {journal} {\bibinfo  {journal} {Journal of Statistical Physics}\ }\textbf
  {\bibinfo {volume} {160}},\ \bibinfo {pages} {794} (\bibinfo {year}
  {2015})}\BibitemShut {NoStop}%
\bibitem [{\citenamefont {Quastel}\ and\ \citenamefont
  {Spohn}(2015)}]{quastel_spohn_2015_JStatPhys160_965}%
  \BibitemOpen
  \bibfield  {author} {\bibinfo {author} {\bibfnamefont {J.}~\bibnamefont
  {Quastel}}\ and\ \bibinfo {author} {\bibfnamefont {H.}~\bibnamefont
  {Spohn}},\ }\href {https://doi.org/10.1007/s10955-015-1250-9} {\bibfield
  {journal} {\bibinfo  {journal} {Journal of Statistical Physics}\ }\textbf
  {\bibinfo {volume} {160}},\ \bibinfo {pages} {965} (\bibinfo {year}
  {2015})}\BibitemShut {NoStop}%
\bibitem [{\citenamefont {Bouchaud}\ \emph {et~al.}(1995)\citenamefont
  {Bouchaud}, \citenamefont {Mézard},\ and\ \citenamefont
  {Parisi}}]{bouchaud_scaling_1995}%
  \BibitemOpen
  \bibfield  {author} {\bibinfo {author} {\bibfnamefont {J.~P.}\ \bibnamefont
  {Bouchaud}}, \bibinfo {author} {\bibfnamefont {M.}~\bibnamefont {Mézard}},\
  and\ \bibinfo {author} {\bibfnamefont {G.}~\bibnamefont {Parisi}},\ }\href
  {https://doi.org/10.1103/PhysRevE.52.3656} {\bibfield  {journal} {\bibinfo
  {journal} {Phys. Rev. E}\ }\textbf {\bibinfo {volume} {52}},\ \bibinfo
  {pages} {3656} (\bibinfo {year} {1995})}\BibitemShut {NoStop}%
\bibitem [{\citenamefont {Korshunov}\ and\ \citenamefont
  {Dotsenko}(1998)}]{korshunov_fluctuations_1998}%
  \BibitemOpen
  \bibfield  {author} {\bibinfo {author} {\bibfnamefont {S.~E.}\ \bibnamefont
  {Korshunov}}\ and\ \bibinfo {author} {\bibfnamefont {V.~S.}\ \bibnamefont
  {Dotsenko}},\ }\href {https://doi.org/10.1088/0305-4470/31/11/009} {\bibfield
   {journal} {\bibinfo  {journal} {J. Phys. A: Math. Gen.}\ }\textbf {\bibinfo
  {volume} {31}},\ \bibinfo {pages} {2591} (\bibinfo {year}
  {1998})}\BibitemShut {NoStop}%
\bibitem [{\citenamefont {Agoritsas}\ \emph
  {et~al.}(2012{\natexlab{b}})\citenamefont {Agoritsas}, \citenamefont
  {Bustingorry}, \citenamefont {Lecomte}, \citenamefont {Schehr},\ and\
  \citenamefont {Giamarchi}}]{agoritsas-2012-FHHpenta}%
  \BibitemOpen
  \bibfield  {author} {\bibinfo {author} {\bibfnamefont {E.}~\bibnamefont
  {Agoritsas}}, \bibinfo {author} {\bibfnamefont {S.}~\bibnamefont
  {Bustingorry}}, \bibinfo {author} {\bibfnamefont {V.}~\bibnamefont
  {Lecomte}}, \bibinfo {author} {\bibfnamefont {G.}~\bibnamefont {Schehr}},\
  and\ \bibinfo {author} {\bibfnamefont {T.}~\bibnamefont {Giamarchi}},\ }\href
  {https://doi.org/10.1103/PhysRevE.86.031144} {\bibfield  {journal} {\bibinfo
  {journal} {Phys. Rev. E}\ }\textbf {\bibinfo {volume} {86}},\ \bibinfo
  {pages} {031144} (\bibinfo {year} {2012}{\natexlab{b}})}\BibitemShut
  {NoStop}%
\bibitem [{\citenamefont
  {Dotsenko}(2016)}]{dotsenko_2016_JStatMech2016_123304}%
  \BibitemOpen
  \bibfield  {author} {\bibinfo {author} {\bibfnamefont {V.}~\bibnamefont
  {Dotsenko}},\ }\href {https://doi.org/10.1088/1742-5468/aa4e5e} {\bibfield
  {journal} {\bibinfo  {journal} {J. Stat. Mech.: Th. Exp.}\ }\textbf {\bibinfo
  {volume} {2016}},\ \bibinfo {pages} {123304} (\bibinfo {year}
  {2016})}\BibitemShut {NoStop}%
\bibitem [{\citenamefont {Agoritsas}\ and\ \citenamefont
  {Lecomte}(2017)}]{agoritsas_lecomte_2017_JPhysA50_104001}%
  \BibitemOpen
  \bibfield  {author} {\bibinfo {author} {\bibfnamefont {E.}~\bibnamefont
  {Agoritsas}}\ and\ \bibinfo {author} {\bibfnamefont {V.}~\bibnamefont
  {Lecomte}},\ }\href {https://doi.org/10.1088/1751-8121/aa5753} {\bibfield
  {journal} {\bibinfo  {journal} {J. Phys. A: Math. Theor.}\ }\textbf {\bibinfo
  {volume} {50}},\ \bibinfo {pages} {104001} (\bibinfo {year}
  {2017})}\BibitemShut {NoStop}%
\bibitem [{\citenamefont {Mathey}\ \emph {et~al.}(2017)\citenamefont {Mathey},
  \citenamefont {Agoritsas}, \citenamefont {Kloss}, \citenamefont {Lecomte},\
  and\ \citenamefont {Canet}}]{mathey_agoritsas_2017_PhysRevE95_032117}%
  \BibitemOpen
  \bibfield  {author} {\bibinfo {author} {\bibfnamefont {S.}~\bibnamefont
  {Mathey}}, \bibinfo {author} {\bibfnamefont {E.}~\bibnamefont {Agoritsas}},
  \bibinfo {author} {\bibfnamefont {T.}~\bibnamefont {Kloss}}, \bibinfo
  {author} {\bibfnamefont {V.}~\bibnamefont {Lecomte}},\ and\ \bibinfo {author}
  {\bibfnamefont {L.}~\bibnamefont {Canet}},\ }\href
  {https://doi.org/10.1103/PhysRevE.95.032117} {\bibfield  {journal} {\bibinfo
  {journal} {Phys. Rev. E}\ }\textbf {\bibinfo {volume} {95}},\ \bibinfo
  {pages} {032117} (\bibinfo {year} {2017})}\BibitemShut {NoStop}%
\bibitem [{\citenamefont
  {Dotsenko}(2018)}]{dotsenko_2018_JStatMech2018_083302}%
  \BibitemOpen
  \bibfield  {author} {\bibinfo {author} {\bibfnamefont {V.}~\bibnamefont
  {Dotsenko}},\ }\href {https://doi.org/10.1088%2F1742-5468%2Faad6c8}
  {\bibfield  {journal} {\bibinfo  {journal} {J. Stat. Mech.: Th. Exp.}\
  }\textbf {\bibinfo {volume} {2018}},\ \bibinfo {pages} {083302} (\bibinfo
  {year} {2018})}\BibitemShut {NoStop}%
\bibitem [{\citenamefont {Takeuchi}\ \emph {et~al.}(2011)\citenamefont
  {Takeuchi}, \citenamefont {Sano}, \citenamefont {Sasamoto},\ and\
  \citenamefont {Spohn}}]{takeuchi_growing_2011}%
  \BibitemOpen
  \bibfield  {author} {\bibinfo {author} {\bibfnamefont {K.~A.}\ \bibnamefont
  {Takeuchi}}, \bibinfo {author} {\bibfnamefont {M.}~\bibnamefont {Sano}},
  \bibinfo {author} {\bibfnamefont {T.}~\bibnamefont {Sasamoto}},\ and\
  \bibinfo {author} {\bibfnamefont {H.}~\bibnamefont {Spohn}},\ }\href
  {https://doi.org/10.1038/srep00034} {\bibfield  {journal} {\bibinfo
  {journal} {Scientific Reports}\ }\textbf {\bibinfo {volume} {1}},\ \bibinfo
  {pages} {34} (\bibinfo {year} {2011})}\BibitemShut {NoStop}%
\bibitem [{\citenamefont {Takeuchi}\ and\ \citenamefont
  {Sano}(2012)}]{takeuchi_evidence_2012}%
  \BibitemOpen
  \bibfield  {author} {\bibinfo {author} {\bibfnamefont {K.~A.}\ \bibnamefont
  {Takeuchi}}\ and\ \bibinfo {author} {\bibfnamefont {M.}~\bibnamefont
  {Sano}},\ }\href {https://doi.org/10.1007/s10955-012-0503-0} {\bibfield
  {journal} {\bibinfo  {journal} {Journal of Statistical Physics}\ }\textbf
  {\bibinfo {volume} {147}},\ \bibinfo {pages} {853} (\bibinfo {year}
  {2012})}\BibitemShut {NoStop}%
\bibitem [{\citenamefont
  {Takeuchi}(2018{\natexlab{a}})}]{takeuchi_appetizer_2018}%
  \BibitemOpen
  \bibfield  {author} {\bibinfo {author} {\bibfnamefont {K.~A.}\ \bibnamefont
  {Takeuchi}},\ }\href
  {http://www.sciencedirect.com/science/article/pii/S0378437118303170}
  {\bibfield  {journal} {\bibinfo  {journal} {Physica A}\ }\textbf {\bibinfo
  {volume} {504}},\ \bibinfo {pages} {77} (\bibinfo {year}
  {2018}{\natexlab{a}})}\BibitemShut {NoStop}%
\bibitem [{\citenamefont {Gotoh}\ and\ \citenamefont
  {Kraichnan}(1998)}]{gotoh_steady-state_1998}%
  \BibitemOpen
  \bibfield  {author} {\bibinfo {author} {\bibfnamefont {T.}~\bibnamefont
  {Gotoh}}\ and\ \bibinfo {author} {\bibfnamefont {R.~H.}\ \bibnamefont
  {Kraichnan}},\ }\href {https://doi.org/10.1063/1.869807} {\bibfield
  {journal} {\bibinfo  {journal} {Physics of Fluids}\ }\textbf {\bibinfo
  {volume} {10}},\ \bibinfo {pages} {2859} (\bibinfo {year}
  {1998})}\BibitemShut {NoStop}%
\bibitem [{\citenamefont {Bec}\ and\ \citenamefont
  {Khanin}(2007)}]{bec_burgers_2007}%
  \BibitemOpen
  \bibfield  {author} {\bibinfo {author} {\bibfnamefont {J.}~\bibnamefont
  {Bec}}\ and\ \bibinfo {author} {\bibfnamefont {K.}~\bibnamefont {Khanin}},\
  }\href {https://doi.org/10.1016/j.physrep.2007.04.002} {\bibfield  {journal}
  {\bibinfo  {journal} {Physics Reports}\ }\textbf {\bibinfo {volume} {447}},\
  \bibinfo {pages} {1} (\bibinfo {year} {2007})}\BibitemShut {NoStop}%
\bibitem [{\citenamefont {Edwards}\ and\ \citenamefont
  {Wilkinson}(1982)}]{edwards_wilkinson}%
  \BibitemOpen
  \bibfield  {author} {\bibinfo {author} {\bibfnamefont {S.~F.}\ \bibnamefont
  {Edwards}}\ and\ \bibinfo {author} {\bibfnamefont {D.~R.}\ \bibnamefont
  {Wilkinson}},\ }\href {https://doi.org/10.1098/rspa.1982.0056} {\bibfield
  {journal} {\bibinfo  {journal} {Proceedings of the Royal Society A}\ }\textbf
  {\bibinfo {volume} {381}},\ \bibinfo {pages} {17} (\bibinfo {year}
  {1982})}\BibitemShut {NoStop}%
\bibitem [{\citenamefont {Kolton}\ \emph {et~al.}(2005)\citenamefont {Kolton},
  \citenamefont {Rosso},\ and\ \citenamefont
  {Giamarchi}}]{kolton_2005_PhysRevLett94_047002}%
  \BibitemOpen
  \bibfield  {author} {\bibinfo {author} {\bibfnamefont {A.~B.}\ \bibnamefont
  {Kolton}}, \bibinfo {author} {\bibfnamefont {A.}~\bibnamefont {Rosso}},\ and\
  \bibinfo {author} {\bibfnamefont {T.}~\bibnamefont {Giamarchi}},\ }\href
  {https://doi.org/10.1103/PhysRevLett.94.047002} {\bibfield  {journal}
  {\bibinfo  {journal} {Phys. Rev. Lett.}\ }\textbf {\bibinfo {volume} {94}},\
  \bibinfo {pages} {047002} (\bibinfo {year} {2005})}\BibitemShut {NoStop}%
\bibitem [{\citenamefont {Ferrero}\ \emph
  {et~al.}(2013{\natexlab{a}})\citenamefont {Ferrero}, \citenamefont
  {Bustingorry}, \citenamefont {Kolton},\ and\ \citenamefont
  {Rosso}}]{ferrero_2013_ComptesRendusPhys14_641}%
  \BibitemOpen
  \bibfield  {author} {\bibinfo {author} {\bibfnamefont {E.~E.}\ \bibnamefont
  {Ferrero}}, \bibinfo {author} {\bibfnamefont {S.}~\bibnamefont
  {Bustingorry}}, \bibinfo {author} {\bibfnamefont {A.~B.}\ \bibnamefont
  {Kolton}},\ and\ \bibinfo {author} {\bibfnamefont {A.}~\bibnamefont
  {Rosso}},\ }\href {https://doi.org/10.1016/j.crhy.2013.08.002} {\bibfield
  {journal} {\bibinfo  {journal} {C. R. Physique}\ }\textbf {\bibinfo {volume}
  {14}},\ \bibinfo {pages} {641} (\bibinfo {year}
  {2013}{\natexlab{a}})}\BibitemShut {NoStop}%
\bibitem [{\citenamefont {Ferrero}\ \emph
  {et~al.}(2013{\natexlab{b}})\citenamefont {Ferrero}, \citenamefont
  {Bustingorry},\ and\ \citenamefont {Kolton}}]{Ferrero2013nonsteady}%
  \BibitemOpen
  \bibfield  {author} {\bibinfo {author} {\bibfnamefont {E.~E.}\ \bibnamefont
  {Ferrero}}, \bibinfo {author} {\bibfnamefont {S.}~\bibnamefont
  {Bustingorry}},\ and\ \bibinfo {author} {\bibfnamefont {A.~B.}\ \bibnamefont
  {Kolton}},\ }\href {https://doi.org/10.1103/PhysRevE.87.032122} {\bibfield
  {journal} {\bibinfo  {journal} {Phys. Rev. E}\ }\textbf {\bibinfo {volume}
  {87}},\ \bibinfo {pages} {032122} (\bibinfo {year}
  {2013}{\natexlab{b}})}\BibitemShut {NoStop}%
\bibitem [{\citenamefont {Salmon}\ \emph {et~al.}(2011)\citenamefont {Salmon},
  \citenamefont {Moraes}, \citenamefont {Dror},\ and\ \citenamefont
  {Shaw}}]{salmon2011parallel}%
  \BibitemOpen
  \bibfield  {author} {\bibinfo {author} {\bibfnamefont {J.~K.}\ \bibnamefont
  {Salmon}}, \bibinfo {author} {\bibfnamefont {M.~A.}\ \bibnamefont {Moraes}},
  \bibinfo {author} {\bibfnamefont {R.~O.}\ \bibnamefont {Dror}},\ and\
  \bibinfo {author} {\bibfnamefont {D.~E.}\ \bibnamefont {Shaw}},\ }in\
  \href@noop {} {\emph {\bibinfo {booktitle} {Proceedings of 2011 International
  Conference for High Performance Computing, Networking, Storage and
  Analysis}}}\ (\bibinfo {year} {2011})\ pp.\ \bibinfo {pages}
  {1--12}\BibitemShut {NoStop}%
\bibitem [{\citenamefont {Kardar}(1985)}]{kardar_1985_PhysRevLett55_2923}%
  \BibitemOpen
  \bibfield  {author} {\bibinfo {author} {\bibfnamefont {M.}~\bibnamefont
  {Kardar}},\ }\href {https://doi.org/10.1103/PhysRevLett.55.2923} {\bibfield
  {journal} {\bibinfo  {journal} {Physical Review Letters}\ }\textbf {\bibinfo
  {volume} {55}},\ \bibinfo {pages} {2923} (\bibinfo {year} {1985})},\ \bibinfo
  {note} {publisher: American Physical Society}\BibitemShut {NoStop}%
\bibitem [{\citenamefont {Kardar}\ and\ \citenamefont
  {Zhang}(1987)}]{kardar_1987_PhysRevLett58_2087}%
  \BibitemOpen
  \bibfield  {author} {\bibinfo {author} {\bibfnamefont {M.}~\bibnamefont
  {Kardar}}\ and\ \bibinfo {author} {\bibfnamefont {Y.-C.}\ \bibnamefont
  {Zhang}},\ }\href {https://doi.org/10.1103/PhysRevLett.58.2087} {\bibfield
  {journal} {\bibinfo  {journal} {Phys. Rev. Lett.}\ }\textbf {\bibinfo
  {volume} {58}},\ \bibinfo {pages} {2087} (\bibinfo {year}
  {1987})}\BibitemShut {NoStop}%
\bibitem [{\citenamefont {Caballero}\ \emph {et~al.}(2020)\citenamefont
  {Caballero}, \citenamefont {Agoritsas}, \citenamefont {Lecomte},\ and\
  \citenamefont {Giamarchi}}]{caballero_2020_PhysRevB102_104204}%
  \BibitemOpen
  \bibfield  {author} {\bibinfo {author} {\bibfnamefont {N.}~\bibnamefont
  {Caballero}}, \bibinfo {author} {\bibfnamefont {E.}~\bibnamefont
  {Agoritsas}}, \bibinfo {author} {\bibfnamefont {V.}~\bibnamefont {Lecomte}},\
  and\ \bibinfo {author} {\bibfnamefont {T.}~\bibnamefont {Giamarchi}},\ }\href
  {https://doi.org/10.1103/PhysRevB.102.104204} {\bibfield  {journal} {\bibinfo
   {journal} {Phys. Rev. B}\ }\textbf {\bibinfo {volume} {102}},\ \bibinfo
  {pages} {104204} (\bibinfo {year} {2020})}\BibitemShut {NoStop}%
\bibitem [{\citenamefont {Bustingorry}\ \emph {et~al.}(2021)\citenamefont
  {Bustingorry}, \citenamefont {Guyonnet}, \citenamefont {Paruch},\ and\
  \citenamefont
  {Agoritsas}}]{bustingorry_agoritsas_2021_JPhysCondensMatter33_345001}%
  \BibitemOpen
  \bibfield  {author} {\bibinfo {author} {\bibfnamefont {S.}~\bibnamefont
  {Bustingorry}}, \bibinfo {author} {\bibfnamefont {J.}~\bibnamefont
  {Guyonnet}}, \bibinfo {author} {\bibfnamefont {P.}~\bibnamefont {Paruch}},\
  and\ \bibinfo {author} {\bibfnamefont {E.}~\bibnamefont {Agoritsas}},\ }\href
  {https://doi.org/10.1088/1361-648x/ac0b20} {\bibfield  {journal} {\bibinfo
  {journal} {Journal of Physics: Condensed Matter}\ }\textbf {\bibinfo {volume}
  {33}},\ \bibinfo {pages} {345001} (\bibinfo {year} {2021})}\BibitemShut
  {NoStop}%
\bibitem [{Note1()}]{Note1}%
  \BibitemOpen
  \bibinfo {note} {In Appendix~\ref {appendix-disorder-Gaussian-distribution},
  we compare our results for two very different disorder distribution,
  generated from potentials which have bounded and unbounded support. We show
  that ${B_\protect \text {dis}(r,T)}$ presents the same exponent $\zeta
  _\protect \text {dis}$, providing compelling evidence for the universality of
  the power-law regime characterized by this exponent.}\BibitemShut {Stop}%
\bibitem [{\citenamefont {Larkin}(1970)}]{Larkin_model_1970-SovPhysJETP31_784}%
  \BibitemOpen
  \bibfield  {author} {\bibinfo {author} {\bibfnamefont {A.~I.}\ \bibnamefont
  {Larkin}},\ }\href@noop {} {\bibfield  {journal} {\bibinfo  {journal} {Sov.
  Phys. JETP}\ }\textbf {\bibinfo {volume} {31}},\ \bibinfo {pages} {784}
  (\bibinfo {year} {1970})}\BibitemShut {NoStop}%
\bibitem [{Note2()}]{Note2}%
  \BibitemOpen
  \bibinfo {note} {See Eqs.~(49), (97) in \cite
  {agoritsas_2012_FHHtri-analytics}}\BibitemShut {NoStop}%
\bibitem [{Note3()}]{Note3}%
  \BibitemOpen
  \bibinfo {note} {The high-temperature version of this scaling choice was
  reported in Table~(2a) of \cite {agoritsas_lecomte_2017_JPhysA50_104001}, but
  had an unpleasant ill-defined limit at asymptotically \protect \emph {large}
  lengthscales for both ${T \to 0}$ and ${\xi \to 0}$. We argue here that it is
  instead relevant for \protect \emph {short} lengthscales.}\BibitemShut
  {Stop}%
\bibitem [{Note4()}]{Note4}%
  \BibitemOpen
  \bibinfo {note} {See Eq.~(99) in \cite {agoritsas_2012_FHHtri-analytics},
  where the DP `time' ${t_{\protect \text {sat}}}$ can be identified with our
  crossover $r_0$.}\BibitemShut {Stop}%
\bibitem [{Note5()}]{Note5}%
  \BibitemOpen
  \bibinfo {note} {At low $T$, ${B_\protect \text {th}(r)}$ intersects
  ${B_\protect \text {dis}(r)}$ at a crossover ${r_1<r_0}$. We can evaluate
  $r_1$ at ${T \ll T_c}$ from ${T r_1/c \equiv A_1 (r_1/r_0)^{2 \zeta _\protect
  \text {dis}}}$, which yields ${r_1^{1-2(1-\zeta _\protect \text {dis})} =
  \protect \frac {T \protect \tmspace +\thinmuskip {.1667em} (T/f)^{4/3}}{c \xi
  ^{4/3} D^{2/3}} r_0^{2(1-\zeta _\protect \text {dis})}}$ with ${T/f \sim
  T_c}$}\BibitemShut {NoStop}%
\bibitem [{\citenamefont {Huse}\ and\ \citenamefont
  {Henley}(1985)}]{huse1985pinning}%
  \BibitemOpen
  \bibfield  {author} {\bibinfo {author} {\bibfnamefont {D.~A.}\ \bibnamefont
  {Huse}}\ and\ \bibinfo {author} {\bibfnamefont {C.~L.}\ \bibnamefont
  {Henley}},\ }\href {https://doi.org/10.1103/PhysRevLett.54.2708} {\bibfield
  {journal} {\bibinfo  {journal} {Phys. Rev. Lett.}\ }\textbf {\bibinfo
  {volume} {54}},\ \bibinfo {pages} {2708} (\bibinfo {year}
  {1985})}\BibitemShut {NoStop}%
\bibitem [{Note6()}]{Note6}%
  \BibitemOpen
  \bibinfo {note} {See Eqs.~(B.12)-(B.13) in \cite
  {agoritsas_2012_FHHtri-analytics}, derived from the statistical tilt
  symmetry. Note that this relation is valid under the condition that
  ${u(0)=0}$ for any given disorder realization $V_p$; since our numerical
  approach does not impose such a boundary condition, we cannot directly
  evaluate $B_\protect \text {dis}(r)$ from such an identity.}\BibitemShut
  {Stop}%
\bibitem [{Note7()}]{Note7}%
  \BibitemOpen
  \bibinfo {note} {See Appendix B in Ref.~\cite
  {agoritsas_2012_FHHtri-analytics}.}\BibitemShut {Stop}%
\bibitem [{Note8()}]{Note8}%
  \BibitemOpen
  \bibinfo {note} {See Eqs.~(56), (126) in \cite
  {agoritsas_2010_PhysRevB_82_184207}, or Eqs.~(6.21), (6.26) in \cite
  {phdthesis_Agoritsas2013}.}\BibitemShut {Stop}%
\bibitem [{Note9()}]{Note9}%
  \BibitemOpen
  \bibinfo {note} {See Eq.~(20) in \cite {korshunov_2013_JETP117_570} for the
  second-order correction at finite temperature, leading to the conjecture
  stated in Eq.~(21).}\BibitemShut {Stop}%
\bibitem [{Note10()}]{Note10}%
  \BibitemOpen
  \bibinfo {note} {See Sec.~VI in \cite {agoritsas_2012_FHHtri-analytics} and
  the summary in Fig.~7.1 in \cite {phdthesis_Agoritsas2013}.}\BibitemShut
  {Stop}%
\bibitem [{\citenamefont {Hartmann}\ \emph {et~al.}(2020)\citenamefont
  {Hartmann}, \citenamefont {Krajenbrink},\ and\ \citenamefont
  {Le~Doussal}}]{hartmann_2020_PhysRevE101_012134}%
  \BibitemOpen
  \bibfield  {author} {\bibinfo {author} {\bibfnamefont {A.~K.}\ \bibnamefont
  {Hartmann}}, \bibinfo {author} {\bibfnamefont {A.}~\bibnamefont
  {Krajenbrink}},\ and\ \bibinfo {author} {\bibfnamefont {P.}~\bibnamefont
  {Le~Doussal}},\ }\href {https://link.aps.org/doi/10.1103/PhysRevE.101.012134}
  {\bibfield  {journal} {\bibinfo  {journal} {Phys. Rev. E}\ }\textbf {\bibinfo
  {volume} {101}},\ \bibinfo {pages} {012134} (\bibinfo {year}
  {2020})}\BibitemShut {NoStop}%
\bibitem [{\citenamefont
  {Takeuchi}(2018{\natexlab{b}})}]{takeuchi_2018_PhysicaA504_77}%
  \BibitemOpen
  \bibfield  {author} {\bibinfo {author} {\bibfnamefont {K.~A.}\ \bibnamefont
  {Takeuchi}},\ }\href
  {http://www.sciencedirect.com/science/article/pii/S0378437118303170}
  {\bibfield  {journal} {\bibinfo  {journal} {Physica A}\ }\textbf {\bibinfo
  {volume} {504}},\ \bibinfo {pages} {77} (\bibinfo {year}
  {2018}{\natexlab{b}})}\BibitemShut {NoStop}%
\bibitem [{Note11()}]{Note11}%
  \BibitemOpen
  \bibinfo {note} {See Sec.~7.2.2 in \cite
  {phdthesis_Agoritsas2013}.}\BibitemShut {Stop}%
\bibitem [{\citenamefont {Cartes}\ \emph {et~al.}(2022)\citenamefont {Cartes},
  \citenamefont {Tirapegui}, \citenamefont {Pandit},\ and\ \citenamefont
  {Brachet}}]{cartes_galerkin-truncated_2022}%
  \BibitemOpen
  \bibfield  {author} {\bibinfo {author} {\bibfnamefont {C.}~\bibnamefont
  {Cartes}}, \bibinfo {author} {\bibfnamefont {E.}~\bibnamefont {Tirapegui}},
  \bibinfo {author} {\bibfnamefont {R.}~\bibnamefont {Pandit}},\ and\ \bibinfo
  {author} {\bibfnamefont {M.}~\bibnamefont {Brachet}},\ }\href
  {https://doi.org/10.1098/rsta.2021.0090} {\bibfield  {journal} {\bibinfo
  {journal} {Philosophical Transactions of the Royal Society A: Mathematical,
  Physical and Engineering Sciences}\ }\textbf {\bibinfo {volume} {380}},\
  \bibinfo {pages} {20210090} (\bibinfo {year} {2022})}\BibitemShut {NoStop}%
\bibitem [{\citenamefont
  {Caballero}(2021)}]{caballero_JSTAT_2021}%
  \BibitemOpen
  \bibfield  {author} {\bibinfo {author} {\bibfnamefont {N.}~\bibnamefont
  {Caballero}},\ }\href {https://dx.doi.org/10.1088/1742-5468/ac2898}
  {\bibfield  {journal} {\bibinfo  {journal} {J. Stat. Mech.: Th. Exp.}\
  }\textbf {\bibinfo {volume} {2021}},\ \bibinfo {pages} {103207} (\bibinfo
  {year} {2021})}\BibitemShut {NoStop}%
\bibitem [{\citenamefont {Domenichini}\ \emph {et~al.}(2019)\citenamefont
  {Domenichini}, \citenamefont {Quinteros}, \citenamefont {Granada},
  \citenamefont {Collin}, \citenamefont {George}, \citenamefont {Curiale},
  \citenamefont {Bustingorry}, \citenamefont {Capeluto},\ and\ \citenamefont
  {Pasquini}}]{domenichini2019transient}%
  \BibitemOpen
  \bibfield  {author} {\bibinfo {author} {\bibfnamefont {P.}~\bibnamefont
  {Domenichini}}, \bibinfo {author} {\bibfnamefont {C.~P.}\ \bibnamefont
  {Quinteros}}, \bibinfo {author} {\bibfnamefont {M.}~\bibnamefont {Granada}},
  \bibinfo {author} {\bibfnamefont {S.}~\bibnamefont {Collin}}, \bibinfo
  {author} {\bibfnamefont {J.-M.}\ \bibnamefont {George}}, \bibinfo {author}
  {\bibfnamefont {J.}~\bibnamefont {Curiale}}, \bibinfo {author} {\bibfnamefont
  {S.}~\bibnamefont {Bustingorry}}, \bibinfo {author} {\bibfnamefont {M.~G.}\
  \bibnamefont {Capeluto}},\ and\ \bibinfo {author} {\bibfnamefont
  {G.}~\bibnamefont {Pasquini}},\ }\href
  {https://doi.org/10.1103/PhysRevB.99.214401} {\bibfield  {journal} {\bibinfo
  {journal} {Phys. Rev. B}\ }\textbf {\bibinfo {volume} {99}},\ \bibinfo
  {pages} {214401} (\bibinfo {year} {2019})}\BibitemShut {NoStop}%
\bibitem [{\citenamefont {Caballero}\ \emph {et~al.}(2018)\citenamefont
  {Caballero}, \citenamefont {Ferrero}, \citenamefont {Kolton}, \citenamefont
  {Curiale}, \citenamefont {Jeudy},\ and\ \citenamefont
  {Bustingorry}}]{caballero2018magnetic}%
  \BibitemOpen
  \bibfield  {author} {\bibinfo {author} {\bibfnamefont {N.~B.}\ \bibnamefont
  {Caballero}}, \bibinfo {author} {\bibfnamefont {E.~E.}\ \bibnamefont
  {Ferrero}}, \bibinfo {author} {\bibfnamefont {A.~B.}\ \bibnamefont {Kolton}},
  \bibinfo {author} {\bibfnamefont {J.}~\bibnamefont {Curiale}}, \bibinfo
  {author} {\bibfnamefont {V.}~\bibnamefont {Jeudy}},\ and\ \bibinfo {author}
  {\bibfnamefont {S.}~\bibnamefont {Bustingorry}},\ }\href
  {https://doi.org/10.1103/PhysRevE.97.062122} {\bibfield  {journal} {\bibinfo
  {journal} {Phys. Rev. E}\ }\textbf {\bibinfo {volume} {97}},\ \bibinfo
  {pages} {062122} (\bibinfo {year} {2018})}\BibitemShut {NoStop}%
\bibitem [{\citenamefont {Agoritsas}(2013)}]{phdthesis_Agoritsas2013}%
  \BibitemOpen
  \bibfield  {author} {\bibinfo {author} {\bibfnamefont {E.}~\bibnamefont
  {Agoritsas}},\ }\emph {\bibinfo {title} {Temperature-dependence of a 1D
  Interface Fluctuations: Role of a Finite Disorder Correlation Length}},\
  \href {http://archive-ouverte.unige.ch/unige:30031} {Ph.D. thesis},\ \bibinfo
   {school} {University of Geneva (Switzerland)} (\bibinfo {year}
  {2013})\BibitemShut {NoStop}%
\end{thebibliography}
%

\end{document}